\documentclass[aps, pra, superscriptaddress, reprint]{revtex4-2}
\setcitestyle{super,open={},close={}}

\usepackage{hyperref, graphicx, xspace}
\hypersetup{hidelinks, colorlinks=true, allcolors=blue}

\usepackage{siunitx, booktabs, multirow, threeparttable, natmove}
\AtBeginDocument{\RenewCommandCopy\qty\SI}
\DeclareSIUnit{\angstrom}{\textup{\AA}}
\DeclareSIUnit{\hartree}{hartree}
\DeclareSIUnit{\bohr}{bohr}

\usepackage[version=4]{mhchem}

\newcommand*{\PySCF}{\textsc{PySCF}\xspace}
\newcommand*{\PySCFforge}{\textsc{PySCF-forge}\xspace}
\newcommand*{\mrh}{\textsc{mrh}\xspace}
\newcommand*{\geomeTRIC}{\textsc{geomeTRIC}\xspace}
\newcommand*{\libxc}{\textsc{libxc}\xspace}
\newcommand*{\libcint}{\textsc{libcint}\xspace}

\usepackage[acronym]{glossaries-extra}
\setabbreviationstyle[acronym]{long-postshort-user}
\glsdisablehyper

\newacronym{ue}{UE}{unsigned error}
\newacronym{ure}{URE}{unsigned relative error}
\newacronym{mue}{MUE}{mean unsigned error}
\newacronym{mse}{MSE}{mean signed error}
\newacronym{rmse}{RMSE}{root-mean-squared error}
\newacronym{zpe}{ZPE}{zero-point energy}
\newacronym{tbe}{TBE}{theoretical best estimate}

\newacronym{pcg}{PCG}{preconditioned conjugate gradient}
\newacronym{csf}{CSF}{configuration state function}

\newacronym{ao}{AO}{atomic orbital}
\newacronym{mo}{MO}{molecular orbital}

\newacronym{lda}{LDA}{local-density approximation}
\newacronym{ga}{GA}{gradient approximation}
\newacronym{mga}{meta-GA}{meta-gradient-approximation}
\newacronym{mgga}{meta-GGA}{meta-generalized-gradient-approximation}
\newacronym{ks-dft}{KS-DFT}{Kohn-Sham density functional theory}

\newacronym{mcpdft}{MC-PDFT}{multiconfiguration pair-density functional theory}
\newacronym{lpdft}{L-PDFT}{linearized pair-density functional theory}

\newacronym{fci}{FCI}{full configuration interaction}

\newacronym{lasscf}{LASSCF}{localized active space self-consistent field}
\newacronym{rasscf}{RASSCF}{restricted active space self-consistent field}
\newacronym{casscf}{CASSCF}{complete active space self-consistent field}
\newacronym{ss-casscf}{SS-CASSCF}{state-specific complete active space self-consistent field theory}
\newacronym{sa-casscf}{SA-CASSCF}{state-averaged complete active space self-consistent field theory}

\newacronym{nevpt2}{NEVPT2}{$n$-electron valence-state second-order perturbation theory}
\newacronym{caspt2}{CASPT2}{complete active space second-order perturbation theory}
\newacronym{ms-caspt2}{MS-CASPT2}{multi-state complete active space second-order perturbation theory}
\newacronym{xms-caspt2}{XMS-CASPT2}{extended multi-state complete active space second-order perturbation theory}

\newacronym{cc2}{CC2}{second-order coupled cluster}
\newacronym{cc3}{CC3}{third-order coupled cluster}
\newacronym{ccsd}{CCSD}{coupled cluster singles doubles}
\newacronym{ccsd(t)}{CCSD(T)}{\gls{ccsd} with perturbative triples correction}

\newacronym{tddft}{TD-DFT}{time-dependent density functional theory}

\usepackage{physics, calrsfs, mathtools, bm, amsfonts, mathrsfs, amssymb}
\DeclareMathAlphabet{\pazocal}{OMS}{zplm}{m}{n}

\newcommand*{\qop}[1]{\ensuremath{\hat{#1}}}
\newcommand*{\lhamop}{\ensuremath{\qop{H}^\mathrm{Lin}}}
\newcommand*{\energy}{\ensuremath{\pazocal{E}}}
\newcommand*{\vnuc}{\ensuremath{V^\mathrm{nuc}}}
\newcommand*{\rdm}{\ensuremath{\gamma}}
\newcommand*{\srdm}{\ensuremath{\rdm}}

\newcommand*{\zod}{\ensuremath{\check{\srdm}}}

\newcommand*{\ot}{\ensuremath{E^{\mathrm{ot}}}}

\newcommand*{\otk}{\epsilon^{\mathrm{ot}}}
\newcommand*{\xck}{\epsilon^{\mathrm{xc}}}

\newcommand*{\vot}{\vb{v}^\mathrm{ot}}
\newcommand*{\vxc}{\vb{v}^\mathrm{xc}}

\newcommand*{\rhop}{\nabla\rho}
\newcommand*{\phip}{\nabla\phi}

\newcommand*{\vrho}{\bm{\rho}}

\newcommand*{\vtrho}{\bm{\tilde{\rho}}}
\newcommand*{\rhoa}{\rho_\uparrow}
\newcommand*{\rhob}{\rho_\downarrow}

\newcommand*{\sigmaaa}{\sigma_{\uparrow\uparrow}}
\newcommand*{\sigmaab}{\sigma_{\uparrow\downarrow}}
\newcommand*{\sigmabb}{\sigma_{\downarrow\downarrow}}
\newcommand*{\taup}{\nabla\tau}
\newcommand*{\taua}{\tau_{\uparrow}}
\newcommand*{\taub}{\tau_{\downarrow}}
\newcommand*{\tzeta}{\zeta_\mathrm{t}}

\DeclareMathOperator{\jacobian}{\vb{J}}
\DeclareMathOperator{\hess}{\vb{H}}

\usepackage{cleveref}

\begin{document}

\title{MC-PDFT Nuclear Gradients and L-PDFT Energies with Meta and Hybrid Meta On-Top Functionals for Ground- and Excited-State Geometry Optimization and Vertical Excitation Energies}

\author{Matthew R. Hennefarth}
\author{Younghwan Kim}
\author{Bhavnesh Jangid}
\author{Jacob Wardzala}
\author{Matthew R. Hermes}
\affiliation{Department of Chemistry and Chicago Center for Theoretical Chemistry, University of Chicago, Chicago, IL 60637, USA}

\author{Donald G. Truhlar} \email[corresponding author: ]{truhlar@umn.edu}
\affiliation{Department of Chemistry, Chemical Theory Center, and Minnesota Supercomputing Institute, University of Minnesota, Minneapolis, MN 55455-0431, USA}

\author{Laura Gagliardi} \email[corresponding author: ]{lgagliardi@uchicago.edu} 
\affiliation{Department of Chemistry and Chicago Center for Theoretical Chemistry, University of Chicago, Chicago, IL 60637, USA}
\affiliation{Pritzker School of Molecular Engineering, University of Chicago, Chicago, IL 60637, USA}

\date{August 18, 2025}

\begin{abstract}
    \Gls{mcpdft} is a post-MCSCF multireference electronic-structure method that explicitly models strong electron correlation, and \gls{lpdft} is a recently developed multi-state extension that can accurately model conical intersections and locally-avoided crossings. Because \gls{mcpdft} and \gls{lpdft} rely on an on-top energy functional, their accuracy depends on the quality of the on-top functional used. Recent work has introduced translated \gls{mga} on-top functionals, and specifically the MC23 hybrid \gls{mga} on-top functional, which is the first on-top functional specifically optimized for \gls{mcpdft}. Here we report the derivation and implementation of analytic nuclear gradients for \gls{mcpdft} calculations using \gls{mga} and hybrid \gls{mga} on-top functionals. This development also enables analytic nuclear gradients for the widely successful tPBE0 hybrid on-top functional. Because \gls{mcpdft} nuclear-gradient calculations involve the derivative of the on-top functional, this development also enables the use of \gls{mga} on-top functionals in \gls{lpdft} single-point energy calculations. We use the new capabilities to test MC23 for ground-state geometries, excited-state geometries, and vertical excitation energies of \textit{s-trans}-butadiene and benzophenone as well as to test MC23, another hybrid \gls{mga}, and seven other \gls{mga} on-top functionals for 441 vertical excitation energies. We find MC23 performs the best of all nine meta and hybrid meta functionals for vertical excitation energies and is comparable in accuracy to tPBE0 and to the NEVPT2 multireference wave function method. Additionally, we directly compare our \gls{mcpdft} vertical excitation results to previously computed TD-DFT values and find that \gls{mcpdft} outperforms even the best performing Kohn-Sham density functional.
\end{abstract}

\glsresetall

\maketitle

\newpage

\section{Introduction}

A long-standing goal of the computational chemistry community is the ability to accurately and efficiently model not only ground-state wave functions and energies of molecular systems but also the excited-state wave functions and energies. Although \gls{ks-dft} \cite{KohnSelfconsistent1965} has excelled for a variety of computational objectives, it is a single reference method; and with currently available functionals, it is less accurate for targets with strong electron correlation -- such as many electronically excited states, transition metal complexes, and bond-stretched molecules -- than for systems whose electronic structure is well represented by a single Slater determinant. This motivates going beyond conventional \gls{ks-dft} to model such processes.

Active-space methods such as \gls{casscf} theory \cite{RoosComplete1980, RoosComplete1987, RuedenbergMCSCF1979} or \gls{lasscf} \cite{HermesMulticonfigurational2019, HermesVariational2020} theory expand the wave function in a linear combination of Slater determinants and can provide qualitatively accurate wave functions even for strongly correlated systems. However, they are not quantitatively accurate because they fail to capture electron correlation external to the active space, which, for a good choice of active space, is mainly dynamic correlation. Multireference perturbation theories such as \gls{caspt2} \cite{AnderssonSecondorder1990, AnderssonSecondorder1992} or \gls{nevpt2} \cite{AngeliIntroduction2001} are widely employed to capture the missing electron correlation external to the active space; however, these methods are computationally expensive and can suffer from slow convergence.

\Gls{mcpdft} \cite{LiManniMulticonfiguration2014} is an alternative post-SCF method that combines features of density functional theory with wave function theory, and it is able to capture the missing dynamic electron correlation. Starting with a multiconfiguration reference wave function, such as from \gls{casscf}, \gls{rasscf} \cite{MalmqvistRestricted1990, ZhouMulticonfiguration2019}, the separated-pair approximation (SP), \cite{OdohSP2016} or \gls{lasscf} \cite{PandharkarLocalized2021} -- \gls{mcpdft} computes the energy using an expression that includes an on-top energy functional. Because \gls{mcpdft} computes the energy in a one-shot manner, it generally adds no significant computational cost over the cost of calculating the reference wave function. Furthermore, it has been shown that \gls{mcpdft} performs similarly to \gls{caspt2} \cite{HoyerMulticonfiguration2016} and \gls{nevpt2} \cite{KingLargescale2022} in predicting vertical excitation energies, and it has been successfully applied to study the nonadiabatic photodynamics of thioformaldehyde \cite{CalioNonadiabatic2022} and azomethane \cite{HennefarthSemiclassical2024}.

As in \gls{ks-dft}, the quality of the calculated energy is highly dependent on the choice of functional. In most work to date, \gls{mcpdft} has used ``translated'' \cite{LiManniMulticonfiguration2014} or ``fully-translated'' \cite{CarlsonMulticonfiguration2015} \gls{ks-dft} \gls{lda} and \gls{ga} functionals. These functionals were not specifically parameterized for \gls{mcpdft}. Recently, a translation scheme for \gls{mga} functionals has been devised, and the hybrid \gls{mga} MC23 on-top functional was introduced and parameterized for \gls{mcpdft} ground-state energies \cite{BaoHybrid2025}.

Analytic nuclear gradients are necessary for geometry optimizations and direct dynamics simulations, but prior to the present work, the newly developed and optimized \gls{mga} functionals were implemented only for \gls{mcpdft} energies. Here, we report the derivation and implementation of analytic nuclear gradients for \gls{mga} functionals and hybrid \gls{mga} functionals (with and without density fitting) within state-specific \gls{mcpdft} (SS-MC-PDFT) and state-averaged \gls{mcpdft} (SA-MC-PDFT). Our derivation and implementation also enable analytic nuclear gradients for hybrid translated \gls{lda} functionals and hybrid translated \gls{ga} functionals (such as the broadly successful tPBE0 \cite{PandharkarNew2020} functional). We first show that the analytic nuclear gradients agree with the numerical gradients for \ce{LiH}. Then we demonstrate the utility of nuclear gradients in predicting ground- and excited-state geometries and adiabatic excitation energies for \textit{s-trans}-butadiene and benzophenone.

We also consider \gls{lpdft}, \cite{HennefarthLinearized2023} which is a multi-state extension of \gls{mcpdft} that can accurately treat potential energy surfaces near conical intersections and locally-avoided crossings. \Gls{lpdft} differs from \gls{mcpdft} in that it is more robust in modeling photochemical and photophysical processes, and it becomes computationally more efficient than \gls{mcpdft} as the number of states within the model space grows because it only requires a single quadrature calculation regardless of the number of states within the model space. Because the \gls{lpdft} Hamiltonian is generated from a first-order Taylor expansion of the \gls{mcpdft} energy functional, it requires the first derivative of the on-top functional with respect to the matrix elements of the one-particle reduced density matix (1-RDM) and two-particle reduced density matrix (2-RDM). These expressions were not previously derived for \gls{mga} on-top functionals), but along with the analytic nuclear gradients for \gls{mcpdft}, we also derive the first derivative of the \gls{mga} on-top functionals with respect to the 1-RDM and 2-RDM, which allows us to perform \gls{lpdft} energy calculations with \gls{mga} functionals. We use this new capability to report the first \gls{lpdft} calculations using \gls{mga} functionals. Using the previously published QUEST wave functions \cite{KingLargescale2022}, we benchmark both \gls{mcpdft} and \gls{lpdft} vertical excitation energies on the QUEST dataset \cite{LoosReference2019, LoosMountaineering2018, LoosMountaineering2020, LoosMountaineering2020a, VerilQUESTDB2021, LoosQuest2020} using a variety of \gls{mga} on-top functionals. 

\section{Theory}

Throughout this manuscript, lower-case roman letters $p,q,r,s,x,y$ indicate general spatial \glspl{mo}, $I$ and $J$ label \gls{casscf} eigenstates within the model space, $M$ and $N$ label \gls{casscf} eigenstates in the complement of the model space (external to the model
space), $\kappa$ is used for \gls{mo} rotations, $\xi$ for a general nuclear coordinate, and $P$ for rotations between different states. Boldfaced variables are tensors (vectors, matrices, etc.). Einstein summation notation is used throughout (repeated indices are summed implicitly). Translated and fully-translated \gls{ks-dft} exchange-correlation functionals have respectively a `t' or `ft' prepended to their name.

\subsection{\Glsentrylong{mcpdft}}
Until recently, the translated and fully translated on-top functionals were restricted to (pure or hybrid) translated and fully-translated \gls{lda} and \gls{ga} exchange-correlation functionals \cite{LiManniMulticonfiguration2014, CarlsonMulticonfiguration2015}. Recently, a translation scheme for \gls{mga} exchange-correlation functionals has been devised \cite{BaoHybrid2025}. The resulting \gls{mga} on-top functionals depend on four arguments: the electron density ($\rho$), the gradient of the electron density ($\rhop$), the kinetic energy density of the electrons ($\tau$), and the on-top pair density ($\Pi$). We introduce the variable $\vrho$ for the collection of these four argument functions:
\begin{equation}
    \vrho = \begin{bmatrix}
        \rho \\ 
        \Pi \\ 
        \rhop \\ 
        \tau
    \end{bmatrix}
\end{equation}
The argument functions can be built from elements of the 1-RDM ($\rdm^p_q$) and 2-RDM ($\rdm^{pr}_{qs}$) as
\begin{equation}
    \label{eq:rho-definition}
    \rho = \phi_p \rdm^p_q \phi^q
\end{equation}
\begin{equation}
    \label{eq:Pi-definition}
    \Pi = \frac{1}{2}\phi_p \phi_r \rdm^{pr}_{qs} \phi^q \phi^s
\end{equation}
\begin{equation}
    \label{eq:rhop-definition}
    \rhop = \pqty{\phip_p \phi^q + \phi_p \phip^q}\rdm^p_q
\end{equation}
\begin{equation}
    \label{eq:tau-definition}
    \tau = \frac{1}{2}\phip_p\cdot\phip^q\rdm^p_q
\end{equation}
where $\phi_p$ and $\phi^q$ are \glspl{mo}.

The \gls{mcpdft} energy for a state with these density matrices is given by \cite{LiManniMulticonfiguration2014,BaoHybrid2025}    
\begin{equation}
    \label{eq:mcpdft-energy}
    \energy^\mathrm{PDFT} = \pqty{h^q_p + \frac{1}{2}\pazocal{J}^q_p\bqty{\bm{\srdm}}}\srdm^p_q + V^\mathrm{nuc} + \ot\bqty{\vrho}
\end{equation}
\begin{equation}
    \pazocal{J}^q_p\bqty{\bm{\srdm}} = g^{qs}_{pr}\rdm^r_s
\end{equation}
where $h^q_p$ and $g^{qs}_{pr}$ are the standard one- and two-electron integrals respectively, $\pazocal{J}^q_p\bqty{\bm{\srdm}}$ is the classical Coulomb interaction, $V^\mathrm{nuc}$ is the nuclear-nuclear repulsion, and $\ot$ is the on-top energy given by an integral over all space of an on-top kernel  $\otk$:
\begin{equation}
    \label{eq:ot-func}
    \ot\bqty{\vrho} = \int \otk\pqty{\vrho\pqty{\vb{r}}} \dd\vb{r}
\end{equation}

In translated functionals, the arguments in $\vrho$ are translated to a set of effective spin-density variables ($\vtrho$) that are then used to evaluate the underlying \gls{ks-dft} exchange-correlation functional. 
\begin{equation}
    \vtrho = \begin{bmatrix}
        \rhoa \\ 
        \rhob \\ 
        \sigmaaa \\ 
        \sigmaab \\ 
        \sigmabb \\ 
        \taua \\ 
        \taub
    \end{bmatrix}
\end{equation}
For \gls{mga} functionals, the translation from $\vrho$ to $\vtrho$ is defined as 
\begin{equation}
    \label{eq:tranlated-mga}
    \vtrho = \begin{bmatrix}
        \frac{\rho}{2}\pqty{1+\tzeta} \\ 
        \frac{\rho}{2}\pqty{1-\tzeta} \\ 
        \frac{\rhop\cdot\rhop}{4}\pqty{1+\tzeta}^2 \\ 
        \frac{\rhop\cdot\rhop}{4}\pqty{1-\tzeta^2} \\ 
        \frac{\rhop\cdot\rhop}{4}\pqty{1-\tzeta}^2 \\ 
        \frac{\tau}{2}\pqty{1+\tzeta} \\ 
        \frac{\tau}{2}\pqty{1-\tzeta} \\
    \end{bmatrix}
\end{equation}
with 
\begin{equation}
    \tzeta = \begin{cases}
        \sqrt{1-R} & R < 1 \\ 
        0 & R \geq 1
    \end{cases}
\end{equation}
and $R$ the ratio between $\Pi$ and ($\rho/2)^2$.
\begin{equation}
    R = \frac{4\Pi}{\rho^2}
\end{equation}

\subsection{\Glsentrylong{lpdft}}
\Gls{lpdft} is a multi-state extension of \gls{mcpdft} that defines the energy operator as the first-order Taylor expansion of its \gls{mcpdft} energy (\cref{eq:mcpdft-energy}) around some zeroth-order density ($\bm{\zod}$) \cite{HennefarthLinearized2023}. 
\begin{equation}
    \energy^\mathrm{Lin} = \energy^\mathrm{PDFT}\bqty{\bm{\zod}} + \eval{\pdv{\energy^\mathrm{PDFT}}{\rdm^p_q}}_{\bm{\zod}}\Delta^p_q + \eval{\pdv{\energy^\mathrm{PDFT}}{\rdm^{pr}_{qs}}}_{\bm{\zod}}\Delta^{pr}_{qs}
\end{equation}
\begin{subequations}
    \begin{align}
        \Delta^p_q &= \srdm^p_q - \zod^p_q \\
        \Delta^{pr}_{qs} &= \srdm^{pr}_{qs} - \zod^{pr}_{qs}
    \end{align}
\end{subequations}
As the \gls{lpdft} energy is linear with respect to both the 1-RDM and the 2-RDM, it can be written as the expectation value of a quantum operator that we call the \gls{lpdft} Hamiltonian.
\begin{equation}
    \energy^\mathrm{Lin}_I = \expval{\lhamop}{I}
\end{equation}
\begin{equation}
    \lhamop = \pqty{h^q_p + \pazocal{J}^q_p\bqty{\bm{\zod}} + V^q_p\bqty{\bm{\zod}}}\qop{E}^p_q + \frac{1}{2}v^{qs}_{pr}\bqty{\bm{\zod}}\qop{e}^{pr}_{qs} + h^\mathrm{const}
\end{equation}
Here, $\qop{E}^p_q$ and $\qop{e}^{pr}_{qs}$ are respectively the singlet excitation operator and the two-electron excitation operator\cite{HelgakerMolecular2000}, and $V^q_p\bqty{\bm{\zod}}$ and $v^{qs}_{pr}\bqty{\bm{\zod}}$ are the one- and two-electron on-top potential terms \cite{SandAnalytic2018}
\begin{subequations}
    \label{eq:on-top-potentials}
    \begin{align}
        V^q_p\bqty{\bm{\zod}} =& \eval{\pdv{\ot}{\rdm^p_q}}_{\bm{\zod}} \\
        v^{qs}_{pr}\bqty{\bm{\zod}} =& 2\eval{\pdv{\ot}{\rdm^{pr}_{qs}}}_{\bm{\zod}}
    \end{align}
\end{subequations}
$h^\mathrm{const}$ is a constant term which only depends on the zeroth-order density.
\begin{equation}
\begin{split}
    h^\mathrm{const} =& \vnuc + \ot \\
    &-\pqty{\frac{1}{2}\pazocal{J}^q_p\bqty{\bm{\zod}} + V^q_p\bqty{\bm{\zod}}}\zod^p_q - \frac{1}{2}v^{qs}_{pr}\bqty{\bm{\zod}}\zod^{pr}_{qs}
\end{split}
\end{equation}
These equations show that the \gls{lpdft} Hamiltonian, and thus its eigenvalues, requires the first derivative of the on-top functional with respect to the elements of the 1-RDM and 2-RDM. These were not previously derived for \gls{mga} functionals. \Cref{sec:mga-on-top-potentials} derives the one- and two-electron on-top potential terms for a translated \gls{mga} functional which allows us to use \gls{mga} functionals with \gls{lpdft}.

\subsection{\Glsentrylong{mga} one- and two-electron on-top potential terms \label{sec:mga-on-top-potentials}}

Both \gls{mcpdft} nuclear gradients and \gls{lpdft} energy calculations require the first derivative of $\ot$ with respect to the elements of the 1-RDM and 2-RDM. Moving the derivative of $\ot$ with respect to the elements of the 1-RDM and 2-RDM inside the integral of \cref{eq:ot-func} and applying the chain rule, we obtain
\begin{subequations}
    \begin{align}
        V^q_p =& \int \vot \cdot \pdv{\vrho}{\rdm^p_q} \dd\vb{r} \\
        v^{qs}_{pr} =& 2\int \vot \cdot \pdv{\vrho}{\rdm^{pr}_{qs}} \dd\vb{r}
    \end{align}
\end{subequations}
where we have defined $\vot$ as 
\begin{equation}
    \vot = \nabla_{\vrho}\otk
\end{equation}
The derivatives of the collection of argument functions with respect to the elements of the 1-RDM and 2-RDM are
\begin{subequations}
    \begin{align}
        \pdv{\vrho}{\rdm^p_q} =& \begin{bmatrix}
            \phi_p\phi^q \\ 
            0 \\ 
            \phip_p\phi^q + \phi_p\phip^q \\ 
            \frac{1}{2}\phip_p\cdot\phip^q
        \end{bmatrix} \\ 
        \pdv{\vrho}{\rdm^{pr}_{qs}} =& \begin{bmatrix}
            0 \\ 
            \frac{1}{2}\phi_p\phi_r\phi^q\phi^s \\ 
            0 \\ 
            0
        \end{bmatrix}
    \end{align}
\end{subequations}

Because $\otk$ is defined as evaluating a Kohn-Sham $\xck$ with the translated effective density variables $\vtrho$, the potential $\vot$ can be evaluated in terms of $\vxc$:
\begin{equation}
    \vot = \vxc\cdot \jacobian^{\vtrho}_{\vrho}
\end{equation}
where $\vxc$ for a \gls{mga} functional is defined as 
\begin{equation}
    \vxc = \begin{bmatrix}
        \pdv{\xck}{\rhoa} & \pdv{\xck}{\rhob} & \pdv{\xck}{\sigmaaa} & \pdv{\xck}{\sigmaab} & \pdv{\xck}{\sigmabb} & \pdv{\xck}{\taua} & \pdv{\xck}{\taub} 
    \end{bmatrix}
\end{equation}
and $\jacobian^{\vtrho}_{\vrho}$ is the Jacobian of the translation from $\vrho$ to $\vtrho$.
\begin{equation}
    \label{eq:mgga-jacobian}
    \jacobian^{\vtrho}_{\vrho} = \begin{bmatrix}
        \pdv{\rhoa}{\rho} & \pdv{\rhoa}{\Pi} & \pdv{\rhoa}{\rhop} & \pdv{\rhoa}{\tau} \\
        \pdv{\rhob}{\rho} & \pdv{\rhob}{\Pi} & \pdv{\rhob}{\rhop} & \pdv{\rhob}{\tau} \\
        \pdv{\sigmaaa}{\rho} & \pdv{\sigmaaa}{\Pi} & \pdv{\sigmaaa}{\rhop} & \pdv{\sigmaaa}{\tau} \\
        \pdv{\sigmaab}{\rho} & \pdv{\sigmaab}{\Pi} & \pdv{\sigmaab}{\rhop} & \pdv{\sigmaab}{\tau} \\
        \pdv{\sigmabb}{\rho} & \pdv{\sigmabb}{\Pi} & \pdv{\sigmabb}{\rhop} & \pdv{\sigmabb}{\tau} \\
        \pdv{\taua}{\rho} & \pdv{\taua}{\Pi} & \pdv{\taua}{\rhop} & \pdv{\taua}{\tau} \\
        \pdv{\taub}{\rho} & \pdv{\taub}{\Pi} & \pdv{\taub}{\rhop} & \pdv{\taub}{\tau}
    \end{bmatrix}
\end{equation}
Although the five upper rows of the Jacobian have been derived previously \cite{SandAnalytic2018, ScottAnalytic2020}, we include their explicit form here for completeness. We start with derivatives of $\rhoa$ and $\rhob$. Using \cref{eq:tranlated-mga}, it can be seen that the only nonzero derivatives of $\rhoa$ and $\rhob$ are with respect to $\rho$ and $\Pi$:
\begin{equation}
    \label{eq:deriv-rhoa-rho}
    \pdv{\rhoa}{\rho} = \frac{1+\tzeta}{2} - R\pdv{\tzeta}{R}
\end{equation}
\begin{equation}
    \label{eq:deriv-rhoa-Pi}
    \pdv{\rhoa}{\Pi} = \frac{2}{\rho}\pdv{\tzeta}{R}
\end{equation}
\begin{equation}
    \label{eq:deriv-rhob-rho}
    \pdv{\rhob}{\rho} = \frac{1-\tzeta}{2} + R\pdv{\tzeta}{R}
\end{equation}
\begin{equation}
    \label{eq:deriv-rhob-Pi}
    \pdv{\rhob}{\Pi} = -\frac{2}{\rho}\pdv{\tzeta}{R}
\end{equation}
where
\begin{equation}
    \pdv{\tzeta}{R} = \begin{cases}
        \frac{-1}{2}\tzeta^{-1} & R < 1 \\ 
        0 & R \geq 1
    \end{cases}
\end{equation}
Due to the stepwise form of $\tzeta$ we do not simplify our equations further. The first derivative of $\sigma_{ab}$ (where $a,b$ can be $\uparrow$ or $\downarrow$) with respect to $\tau$ is $0$. The other, nonzero derivatives are given by
\begin{equation}
    \label{eq:deriv-sigmaaa-rho}
    \pdv{\sigmaaa}{\rho} = \frac{-\abs{\rhop}^2 R\pqty{1+\tzeta}}{\rho}\pdv{\tzeta}{R}
\end{equation}
\begin{equation}
    \label{eq:deriv-sigmaaa-Pi}
    \pdv{\sigmaaa}{\Pi} = \frac{2\abs{\rhop}^2\pqty{1+\tzeta}}{\rho^2}\pdv{\tzeta}{R}
\end{equation}
\begin{equation}
    \label{eq:deriv-sigmaaa-rhop}
    \pdv{\sigmaaa}{\rhop} = \frac{\rhop}{2}\pqty{1+\tzeta}^2
\end{equation}
\begin{equation}
    \label{eq:deriv-sigmaab-rho}
    \pdv{\sigmaab}{\rho} = \frac{\abs{\rhop}^2 R\tzeta}{\rho}\pdv{\tzeta}{R}
\end{equation}
\begin{equation}
    \label{eq:deriv-sigmaab-Pi}
    \pdv{\sigmaab}{\Pi} = \frac{-2\abs{\rhop}^2\tzeta}{\rho^2}\pdv{\tzeta}{R}
\end{equation}

\begin{equation}
    \label{eq:deriv-sigmaab-rhop}
    \pdv{\sigmaab}{\rhop} = \frac{\rhop}{2}\pqty{1-\tzeta^2}
\end{equation}
\begin{equation}
    \label{eq:deriv-sigmabb-rho}
    \pdv{\sigmabb}{\rho} = \frac{\abs{\rhop}^2 R\pqty{1-\tzeta}}{\rho}\pdv{\tzeta}{R}
\end{equation}
\begin{equation}
    \label{eq:deriv-sigmabb-Pi}
    \pdv{\sigmabb}{\Pi} = \frac{-2\abs{\rhop}^2\pqty{1-\tzeta}}{\rho^2}\pdv{\tzeta}{R}
\end{equation}
\begin{equation}
    \label{eq:deriv-sigmabb-rhop}
    \pdv{\sigmabb}{\rhop} = \frac{\rhop}{2}\pqty{1-\tzeta}^2
\end{equation}
Finally, since $\taua$ and $\taub$ do not depend on $\rhop$, those derivatives are zero. The other non-zero derivatives are given by
\begin{equation}
    \label{eq:deriv-taua-rho}
    \pdv{\taua}{\rho} = \frac{-R\tau}{\rho}\pdv{\tzeta}{R}
\end{equation}
\begin{equation}
    \label{eq:deriv-taua-Pi}
    \pdv{\taua}{\Pi} = \frac{2\tau}{\rho^2}\pdv{\tzeta}{R}
\end{equation}
\begin{equation}
    \label{eq:deriv-taua-tau}
    \pdv{\taua}{\tau} = \frac{1+\tzeta}{2}
\end{equation}
\begin{equation}
    \label{eq:deriv-taub-rho}
    \pdv{\taub}{\rho} = \frac{R\tau}{\rho}\pdv{\tzeta}{R}
\end{equation}
\begin{equation}
    \label{eq:deriv-taub-Pi}
    \pdv{\taub}{\Pi} = \frac{-2\tau}{\rho^2}\pdv{\tzeta}{R}
\end{equation}
\begin{equation}
    \label{eq:deriv-taub-tau}
    \pdv{\taub}{\tau} = \frac{1-\tzeta}{2}
\end{equation}
Plugging \cref{eq:deriv-rhoa-rho,eq:deriv-rhoa-Pi,eq:deriv-rhob-rho,eq:deriv-rhob-Pi,eq:deriv-sigmaaa-rho,eq:deriv-sigmaaa-Pi,eq:deriv-sigmaaa-rhop,eq:deriv-sigmaab-rho,eq:deriv-sigmaab-Pi,eq:deriv-sigmaab-rhop,eq:deriv-sigmabb-rho,eq:deriv-sigmabb-Pi,eq:deriv-sigmabb-rhop,eq:deriv-taua-rho,eq:deriv-taua-Pi,eq:deriv-taua-tau,eq:deriv-taub-rho,eq:deriv-taub-Pi,eq:deriv-taub-tau} into \cref{eq:mgga-jacobian} gives the \gls{mga} Jacobian. The last two rows (the derivatives of $\taua$ and $\taub$ with respect to the argument functions) are the new additions in the \gls{mga} Jacobian.

\subsection{\Glsentrylong{mga} \glsentrylong{mcpdft} nuclear gradients}

Because neither SS-\gls{mcpdft} \cite{SandAnalytic2018} nor SA-\gls{mcpdft} \cite{ScottAnalytic2020} variationally optimizes the wave function variables, the analytic nuclear gradients are obtained using Lagrange's method of undetermined multipliers. As the equations for the SS-\gls{mcpdft} nuclear gradients are a special case of the SA-\gls{mcpdft} ones, we need to present only the more general SA-\gls{mcpdft} nuclear gradient equations \cite{ScottAnalytic2020}. Here we first discuss the modifications necessary in the \gls{mcpdft} energy response terms in order to solve for the corresponding Lagrange multipliers. Then, we discuss the modifications necessary to compute the \gls{mcpdft} Hellmann-Feynman contribution.

In general, a \gls{casscf} wave function (state-specific or state-averaged) can be parameterized as 
\begin{equation}
    \ket{I} = e^{\qop{P}^I}e^{\qop{\kappa}}\ket{0}
\end{equation}
where $\qop{\kappa}$ is the \gls{mo}-rotation operator
\begin{equation}
    \label{eq:orbital-rotation-operator}
    \qop{\kappa} = \sum_{p<q} \kappa^q_p\bqty{\qop{\vb{E}} - \qop{\vb{E}}^\dag}^p_q
\end{equation}
and $\qop{P}^I$ is the state-transfer operator for state $\ket{I}$. State rotations can be decomposed into rotations within the model space ($\qop{P}_\parallel$) and rotations out of the model space ($\qop{P}_\perp$).
\begin{equation}
    \qop{P}^I = \qop{P}^I_\perp + \qop{P}^I_\parallel
\end{equation}
\begin{equation}
    \qop{P}^I_\perp = P^I_M\pqty{\ketbra{M}{I} - \ketbra{I}{M}}
\end{equation}
\begin{equation}
    \qop{P}^I_\parallel = P^I_J\pqty{\ketbra{J}{I} - \ketbra{I}{J}}
\end{equation}
In the case of a \gls{ss-casscf} wave function, $\qop{P}^I_\parallel = 0$ as the model space is spanned by a single state ($\ket{I}$).

The Lagrangian for a reference wave function $\ket{I}$ (coming from either \gls{ss-casscf} or \gls{sa-casscf}) is given by 
\begin{equation}
\label{eq:mcpdft-lagrangian}
\begin{split}
    \mathcal{L}^\mathrm{PDFT}_I =& \energy^\mathrm{PDFT}_I + \bar{\bm{\kappa}}\cdot \nabla_{\bm{\kappa}} \energy^\mathrm{SA} + \bar{\vb{P}}_\perp \cdot \nabla_{\vb{P}_\perp} \energy^\mathrm{SA} \\
    &+ \omega_I \bar{\vb{P}}^I_\parallel \cdot \nabla_{\vb{P}^I_\parallel}\energy^\mathrm{CAS}_I
\end{split}
\end{equation}
where $\energy^\mathrm{SA}$ is the average electronic energy within the model space, $\energy^\mathrm{CAS}_I$ is the electronic energy for state $\ket{I}$, $\omega_I$ is the weight of the $I$th root in the state-average, and $\bar{\bm{\kappa}}$, $\bar{\vb{P}}_\perp$, and $\bar{\vb{P}}^I_\parallel$ are the associated Lagrange multipliers for the \gls{mo}-rotation and state-transfer operators respectively. The Lagrange multipliers  $\bar{\bm{\kappa}}$, $\bar{\vb{P}}_\perp$, and $\bar{\vb{P}}^I_\parallel$ are determined by making $\mathcal{L}^\mathrm{PDFT}_I$ stationary with respect to \gls{mo} rotations and state rotations; this requires solving the following system of coupled linear equations.
\begin{equation}
    \label{eq:coupled-perturbed-equations}
    \begin{bmatrix}
        \nabla_{\bm{\kappa}} \energy^\mathrm{PDFT}_I \\ 
        \nabla_{\vb{P}_\perp} \energy^\mathrm{PDFT}_I \\ 
        \nabla_{\vb{P}^I_\parallel} \energy^\mathrm{PDFT}_I
    \end{bmatrix} = - \begin{bmatrix}
        \hess^{\energy^\mathrm{SA}}_{\bm{\kappa}\bm{\kappa}} & \hess^{\energy^\mathrm{SA}}_{\bm{\kappa}\vb{P}_\perp} & \hess^{\energy^\mathrm{CAS}_I}_{\bm{\kappa}\vb{P}^I_\parallel} \\ 
        \hess^{\energy^\mathrm{SA}}_{\vb{P}_\perp\bm{\kappa}} & \hess^{\energy^\mathrm{SA}}_{\vb{P}_\perp\vb{P}_\perp} & \hess^{\energy^\mathrm{CAS}_I}_{\vb{P}_\perp\vb{P}^I_\parallel} \\ 
        0 & 0 & \hess^{\energy^\mathrm{CAS}_I}_{\vb{P}^I_\parallel\vb{P}^I_\parallel}
    \end{bmatrix}
    \begin{bmatrix}
        \bar{\bm{\kappa}} \\ 
        \bar{\vb{P}}_\perp \\ 
        \bar{\vb{P}}^I_\parallel
    \end{bmatrix}
\end{equation}
Here, $\hess^{\energy^\mathrm{SA}}_{\vb{a}\vb{b}}$ denotes the Hessian of $\energy^\mathrm{SA}$ with respect to the $\vb{a}$ and $\vb{b}$ variables.

Because the matrix in \cref{eq:coupled-perturbed-equations} is not symmetric, we cannot use the standard \gls{pcg} method \cite{PressNumerical1992, BernhardssonDirect1999} to solve for all of the Lagrange multipliers in a single step. Instead, we first solve for $\bar{\vb{P}}^I_\parallel$ using
\begin{equation}
    \label{eq:lagrange-first-step}
    \nabla_{\vb{P}^I_\parallel} \energy^\mathrm{PDFT} = -\hess^{\energy^\mathrm{CAS}_I}_{\vb{P}^I_\parallel\vb{P}^I_\parallel} \cdot \bar{\vb{P}}^I_\parallel 
\end{equation}
Then we solve for the remaining Lagrange multipliers  $\bar{\bm{\kappa}}$ and $\bar{\vb{P}}_\perp$ using the standard \gls{pcg} method on
\begin{equation}
    \label{eq:lagrange-second-step}
    \begin{bmatrix}
        \nabla_{\bm{\kappa}} \energy^\mathrm{PDFT} + \hess^{\energy^\mathrm{CAS}_I}_{\bm{\kappa}\vb{P}^I_\parallel} \cdot \bar{\vb{P}}^I_\parallel \\ 
        \nabla_{\vb{P}_\perp} \energy^\mathrm{PDFT} + \hess^{\energy^\mathrm{CAS}_I}_{\vb{P}_\perp\vb{P}^I_\parallel} \cdot \bar{\vb{P}}^I_\parallel
    \end{bmatrix} = - \begin{bmatrix}
        \hess^{\energy^\mathrm{SA}}_{\bm{\kappa}\bm{\kappa}} & \hess^{\energy^\mathrm{SA}}_{\bm{\kappa}\vb{P}_\perp} \\ 
        \hess^{\energy^\mathrm{SA}}_{\vb{P}_\perp\bm{\kappa}} & \hess^{\energy^\mathrm{SA}}_{\vb{P}_\perp\vb{P}_\perp}
    \end{bmatrix}
    \begin{bmatrix}
        \bar{\bm{\kappa}} \\ 
        \bar{\vb{P}}_\perp
    \end{bmatrix}
\end{equation}
In the case of SS-\gls{mcpdft}, there are no model-space rotations, and \cref{eq:coupled-perturbed-equations} can be directly solved using the \gls{pcg} method as all of the $\bar{\vb{P}}^I_\parallel$ Lagrange multipliers are zero.

The response of the \gls{mcpdft} energy to \gls{mo} rotations is known to be 
\begin{equation}
    \pdv{\energy^\mathrm{PDFT}}{\kappa^y_x} = 2\bqty{\mathcal{F}_\mathrm{PDFT} - \mathcal{F}^\dag_\mathrm{PDFT}}^x_y
\end{equation}
where $\mathcal{F}$ is the \gls{mcpdft} generalized Fock matrix \cite{SandAnalytic2018,ScottAnalytic2020}.
\begin{equation}
    \label{eq:mcpdft-generalized-fock}
    \bqty{\mathcal{F}_\mathrm{PDFT}}^x_y = \pqty{h^q_y + \pazocal{J}^q_y\bqty{\bm{\srdm}} + V^q_y\bqty{\bm{\srdm}}} \srdm^x_q + v^{qs}_{yr}\bqty{\bm{\srdm}}\srdm^{xr}_{qs}
\end{equation}
Similarly the response of the \gls{mcpdft} energy with respect to a state rotation is known to be  \cite{ScottAnalytic2020}
\begin{equation}
    \pdv{\energy^\mathrm{PDFT}_I}{P^J_\Lambda} = \delta^I_J \pqty{\mel{I}{\qop{H}^\mathrm{PDFT}_I\qop{Q}_I}{\Lambda} + \mel{\Lambda}{\qop{Q}_I\qop{H}^\mathrm{PDFT}_I}{I}}
\end{equation}
where $\Lambda$ is any \gls{csf}, $\qop{H}^\mathrm{PDFT}_I$ is a operator that depends on the $I$th state through the Coulomb and on-top potential terms, and $\qop{Q}_I$ removes any component along the $I$th root.
\begin{equation}
    \label{eq:mcpdft-effective-ham}
    \qop{H}^\mathrm{PDFT}_I = \pqty{h^q_p + \pazocal{J}^q_p\bqty{\bm{\srdm}} + V^q_p\bqty{\bm{\srdm}}}\qop{E}^p_q + \frac{1}{2}v^{qs}_{pr}\bqty{\bm{\srdm}}\hat{e}^{pr}_{qs}
\end{equation}
\begin{equation}
    \qop{Q}_I = 1 - \ketbra{I}{I}
\end{equation}
The \cref{eq:mcpdft-generalized-fock,eq:mcpdft-effective-ham} show that the one- and two-electron on-top potential terms for \gls{mga} functionals are necessary in order to solve for the Lagrange multipliers, and \cref{sec:mga-on-top-potentials} describes how to calculate these terms.

Having solved for the Lagrange multipliers, the nuclear gradient of $\energy^\mathrm{PDFT}_I$ is given by 
\begin{equation}
    \begin{split}
        \dv{\energy^\mathrm{PDFT}_I}{\xi} =& \pdv{\mathcal{L}^\mathrm{PDFT}_I}{\xi} \\
        =& \pdv{\energy^\mathrm{PDFT}_I}{\xi} + \bar{\bm{\kappa}}\cdot \nabla_{\bm{\kappa}} \pdv{\energy^\mathrm{SA}}{\xi} + \bar{\vb{P}}_\perp \cdot \nabla_{\vb{P}_\perp} \pdv{\energy^\mathrm{SA}}{\xi} \\
        &+ \omega_I \bar{\vb{P}}^I_\parallel \cdot \nabla_{\vb{P}^I_\parallel}\pdv{\energy^\mathrm{CAS}_I}{\xi}
    \end{split}
\end{equation}
All terms except $\pdv{\energy^\mathrm{PDFT}_I}{\xi}$, which is the \gls{mcpdft} Hellmann-Feynman contribution, are unchanged for a \gls{mga} on-top functional. $\pdv{\energy^\mathrm{PDFT}_I}{\xi}$ is given by 
\begin{equation}
\begin{split}
    \pdv{\energy^\mathrm{PDFT}_I}{\xi} =& \pqty{h^q_{p,\xi} + \pazocal{J}^q_{p,\xi}\bqty{\bm{\srdm}}}\srdm^p_q + V^\mathrm{nuc}_{,\xi} + \ot_{,\xi}\bqty{\bm{\rho}_{\bm{\srdm}}} \\
    &- S^q_{p,\xi}\bqty{\mathcal{F}_\mathrm{PDFT}}^p_q
\end{split}
\end{equation}
where $h^q_{p,\xi}$ is the derivative of the one-electron integral, $\pazocal{J}^q_{p,\xi}\bqty{\bm{\srdm}}$ is the derivative of the Coulomb integral, $V^\mathrm{nuc}_{,\xi}$ is the derivative of the nuclear-nuclear repulsion term, $S^q_{p,\xi}$ is the derivative of the \gls{mo} overlap matrix, $\mathcal{F}^p_q$ is the \gls{mcpdft} generalized Fock matrix (\cref{eq:mcpdft-generalized-fock}), and $\ot_{,\xi}\bqty{\bm{\rho}_{\bm{\srdm}}}$ is the derivative of the on-top functional. Note that the comma indicates that the tensor has been differentiated with respect to the following index \cite{MisnerGravitation1973}. The last term is a result of ensuring the \glspl{mo} are orthonormal at all geometries and is sometimes called the ``renormalization'' or ``connection'' term. Only the nuclear derivative of the on-top functional is different for a \gls{mga} on-top functional. Similarly to \cref{sec:mga-on-top-potentials}, we can move the differentiation within the integral and apply the chain-rule to find that 
\begin{equation}
    \ot_{,\xi} = \int \vot \cdot \pdv{\vrho}{\xi} \dd\vb{r} + \Bqty{\otk}_{\vb{r}\in\mathcal{G}} \cdot \pdv{\vb{w}}{\xi}
\end{equation}
where $\mathcal{G}$ is the set of all grid points, $\vb{w}$ are the corresponding quadrature grid weights, and $\Bqty{\otk}_{\vb{r}\in\mathcal{G}}$ is the vector generated by evaluating $\otk$ at all grid points.
\begin{equation}
    \Bqty{\otk}_{\vb{r}\in\mathcal{G}} = \begin{bmatrix}
        \otk\pqty{\vb{r}_1} & \otk\pqty{\vb{r}_2} & \ldots & \otk\pqty{\vb{r}_n}
    \end{bmatrix}
\end{equation}
$\vot$ is evaluated as described in \cref{sec:mga-on-top-potentials} for a \gls{mga} on-top functional. The nuclear derivative of the collected argument functions in $\vrho$ is given by
\begin{equation}
    \pdv{\vrho}{\xi} = \begin{bmatrix}
        2\phi_{p}\srdm^p_q\phi^q_{,\xi} \\ 
        2\phi_p\phi_r\srdm^{pr}_{qs}\phi^q\phi^s_{,\xi} \\ 
        2\pqty{\phip_{p,\lambda}\phi^q + \phip_p\phi^q_{,\xi}}\srdm^p_q \\ 
        \phip_{p,\xi}\cdot\phip^q\srdm^p_q
    \end{bmatrix} + \delta_{\xi}(\vb{r})\begin{bmatrix}
        \nabla \rho \\ 
        \nabla \Pi \\ 
        \hess^\rho_{\vb{r}} \\ 
        \nabla \tau
    \end{bmatrix} \cdot \vb{n}_{\xi}
\end{equation}
where $\delta_{\xi}\pqty{\vb{r}}$ is $1$ if $\vb{r}$ is evaluated at a grid point associated with the atom of the coordinate $\xi$ and is $0$ otherwise; and $\vb{n}_\xi$ is the Cartesian unit vector for the coordinate direction $\xi$. Finally, $\taup$ is given by
\begin{equation}
    \taup = \phip^q \cdot \hess^{\phi_p}_{\vb{r}}\srdm^p_q
\end{equation}
with $\hess^{\phi_p}_{\vb{r}}$ being the Hessian of the $p$th \gls{mo} with respect to the electron coordinate $\vb{r}$.

\subsection{Hybrid \glsentrylong{mcpdft} nuclear gradients}

\Gls{mcpdft} with a  hybrid functional is defined as a weighted average of the \gls{casscf} and non-hybrid \gls{mcpdft} energy where $\lambda$ controls the fraction of \gls{casscf} energy included \cite{PandharkarNew2020}.
\begin{equation}
    \label{eq:hybrid-pdft-energy}
    \energy^\mathrm{Hyb} = \lambda\energy^\mathrm{CAS} + \pqty{1-\lambda}\energy^\mathrm{PDFT}
\end{equation}
For example, the tPBE0 functional sets $\lambda=0.25$, and the MC23 functional sets $\lambda=0.2856$. The resulting hybrid \gls{mcpdft} Lagrangian is 
\begin{equation}
\label{eq:hybrid-mcpdft-lagrangian}
\begin{split}
    \mathcal{L}^\mathrm{Hyb}_I =& \energy^\mathrm{Hyb}_I + \bar{\bm{\kappa}}\cdot \nabla_{\bm{\kappa}} \energy^\mathrm{SA} + \bar{\vb{P}}_\perp \cdot \nabla_{\vb{P}_\perp} \energy^\mathrm{SA} \\
    &+ \omega_I \bar{\vb{P}}^I_\parallel \cdot \nabla_{\vb{P}^I_\parallel}\energy^\mathrm{CAS}_I
\end{split}
\end{equation}
We solve for the Lagrange multipliers in the same way as for the non-hybrid functional (\cref{eq:coupled-perturbed-equations,eq:lagrange-first-step,eq:lagrange-second-step}). For SS-\gls{mcpdft}, the \gls{casscf} energy is stationary with respect to \gls{mo} and state rotations which allows for a further simplification where the non-hybrid SS-\gls{mcpdft} Lagrange multipliers are just scaled by $1-\lambda$. However, in the general SA-\gls{mcpdft}, it requires modified \gls{mcpdft} energy response terms. The resulting hybrid energy response to \gls{mo} rotations is given by
\begin{equation}
    \pdv{\energy^\mathrm{Hyb}}{\kappa^y_x} = 2\bqty{\mathcal{F}_\mathrm{Hyb} - \mathcal{F}^\dag_\mathrm{Hyb}}^x_y
\end{equation}
\begin{equation}
    \mathcal{F}_\mathrm{Hyb} = \lambda\mathcal{F}_\mathrm{CAS} + \pqty{1-\lambda}\mathcal{F}_\mathrm{PDFT}
\end{equation}
where $\mathcal{F}_\mathrm{CAS}$ is the \gls{casscf} generalized Fock matrix.
\begin{equation}
    \bqty{\mathcal{F}_\mathrm{CAS}}^x_y = h^q_y\srdm^x_q + g^{qs}_{yr}\srdm^{xr}_{qs}
\end{equation}
Similarly, the response of the hybrid energy to a state rotation is given by 
\begin{equation}
    \pdv{\energy^\mathrm{Hyb}_I}{P^J_\Lambda} = \delta^I_J\pqty{\mel{I}{\qop{H}^\mathrm{Hyb}_I\qop{Q}_I}{\Lambda} + \mel{\Lambda}{\qop{Q}_I\qop{H}^\mathrm{Hyb}_I}{I}}
\end{equation}
\begin{equation}
    \qop{H}^\mathrm{Hyb}_I = \lambda\qop{H}^\mathrm{el} + \pqty{1-\lambda}\qop{H}^\mathrm{PDFT}_I
\end{equation}
where $\qop{H}^\mathrm{el}$ is the normal electronic Hamiltonian.
\begin{equation}
    \qop{H}^\mathrm{el} = h^q_p \qop{E}^p_q+ \frac{1}{2}g^{qs}_{pr}\qop{e}^{pr}_{qs} + V^\mathrm{nuc}
\end{equation}

Furthermore, once the Lagrange multipliers have been obtained, the \gls{mcpdft} Hellmann-Feynman contribution must be modified as 
\begin{equation}
    \pdv{\energy^\mathrm{Hyb}}{\xi} = \lambda \pdv{\energy^\mathrm{CAS}}{\xi} + \pqty{1-\lambda}\pdv{\energy^\mathrm{PDFT}}{\xi}
\end{equation}
with $\pdv{\energy^\mathrm{CAS}}{\xi}$ the normal \gls{casscf} Hellmann-Feynman contribution.
\begin{equation}
    \pdv{\energy^\mathrm{CAS}}{\xi} = h^q_{p,\xi}\srdm^p_q + \frac{1}{2}g^{qs}_{pr,\xi}\srdm^{pr}_{qs} + V^\mathrm{nuc}_{,\xi} - S^q_{p,\xi}\bqty{\mathcal{F}_\mathrm{CAS}}^p_q
\end{equation}

\subsection{\Glsentrylong{mga} and hybrid nuclear gradients with density fitting}

One of the major bottlenecks in an energy and gradient calculation is computing the four-index electron-repulsion integrals and transforming them from the \gls{ao} basis to the \gls{mo} basis as this scales as $\order{N_\mathrm{basis}^5}$ with $N_\mathrm{basis}$  the number of basis functions. Density fitting procedures -- also known as the resolution of identity -- such as the Cholesky decomposition \cite{BeebeSimplifications1977}, can reduce this scaling by one to two orders of magnitude and are an invaluable technique for practical calculations. Because MC-PDFT density fitting is currently used for the electron-repulsion integrals and not for the two-electron on-top potential terms \cite{scottAnalytic2021}, no modifications are necessary to the nuclear gradients for \gls{mga} or hybrid on-top functionals. 

\section{Computational Methods}

All calculations were performed using \PySCF (version 2.8.0) \cite{sunPySCF2018, sunRecent2020} and \PySCFforge (version 1.0.3, commit \texttt{400b3b6}) \cite{Pyscfforge2025} compiled with \libcint (version 6.1.1) \cite{SunLibcint2015} and \libxc (version 7.0.0) \cite{MarquesLibxc2012, LehtolaRecent2018}. \Gls{mcpdft} \gls{mga} nuclear gradients and \gls{lpdft} \gls{mga} single-point energy calculations are implemented in \PySCFforge, which is available at \url{https://github.com/pyscf/pyscf-forge}.

To validate the analytic gradients, we compared them to numerical gradients for the \ce{LiH} molecule using the aug-cc-pVTZ basis set \cite{DunningGaussian1989, KendallElectron1992} and a (2e, 2o) active space comprising the $\sigma$ and $\sigma^*$ orbitals. 
Calculations were performed both for the ground state using state-specific methods and for the lowest two singlet states using state-averaged methods. All calculations employed the \texttt{csf\_solver} from \mrh (commit \texttt{a88c476d}) \cite{hermesMrh2025}. 
A quadrature grid level of 6 was used, corresponding to 80 or 120 radial points and 770 or 974 angular points for atoms in periods 1 or 2, respectively. Both the hybrid \gls{ga} functional tPBE0 \cite{PandharkarNew2020} and the hybrid \gls{mga} functional MC23 \cite{BaoHybrid2025} were used. 

Geometry optimizations for \textit{s-trans}-butadiene and benzophenone were performed using \geomeTRIC (version 1.0) \cite{wangGeometry2016} with a quadrature grid level of 6. For \textit{s-trans}-butadiene, we used the jul-cc-pVTZ basis set\cite{FellerRole1996, SchuchardtBasis2007, DunningGaussian1989, KendallElectron1992, PapajakConvergent2010} with a (4e, 4o) active space, consisting of the $2\pi$ and $2\pi^*$ orbitals, and we included two $\prescript{1}{}{\mathrm{A}}_\mathrm{g}$ states in the state-averaged manifold. For benzophenone, we used the cc-pVDZ basis set\cite{DunningGaussian1989} with an (8e, 7o) active space consisting of the $n_\mathrm{O}, 2(\pi,\pi^*), \pi_\mathrm{CO}$, and $\pi^*_\mathrm{CO}$ orbitals (\cref{fig:benzophenone-structure}) with two $\prescript{1}{}{\mathrm{A}}$ states and one $\prescript{1}{}{\mathrm{B}}$ state in the state-averaged manifold. We performed the \textit{s-trans}-butadiene calculations by imposing $C_{2\mathrm{h}}$ symmetry and the benzophenone calculations with $C_2$ symmetry. Density fitting \cite{scottAnalytic2021} was employed in the benzophenone calculations.

Nine \gls{mga} and hybrid \gls{mga} functionals were tested for vertical excitations with MC-PDFT and L-PDFT. The tested functionals are
\begin{itemize}
    \item MC23 \cite{BaoHybrid2025}
    \item translated TPSS\cite{TaoClimbing2003, PerdewMetageneralized2004} (tTPSS)
    \item translated TPSSh\cite{StaroverovComparative2003} (tTPSSh, defined as the hybrid, \gls{mga} on-top functional of the translated TPSS exchange-correlation functional with  $\lambda$ set to 0.10 in \cref{eq:hybrid-pdft-energy})
    \item translated SCAN \cite{SunStrongly2015} (tSCAN)
    \item translated r$^2$SCAN \cite{FurnessAccurate2020, FurnessAccurate2020correction} (tr$^2$SCAN)
    \item translated M06-L \cite{ZhaoNew2006} (tM06-L)
    \item translated revM06-L \cite{WangRevised2017} (trevM06-L)
    \item translated MN15-L \cite{YuMN152016} (tMN15-L)
    \item translated $\tau$-HCTH \cite{BoeseNew2002} (t$\tau$-HCTH).
\end{itemize}
Note that MC23 is a hybrid translated \gls{mgga} (a special case of \gls{mga}), tMN5-L is a translated meta-nonseparable-gradient-approximation (another special case of \gls{mga}), tTPSSh is a hybrid translated \gls{mgga}, and the other six are (non-hybrid) translated \glspl{mgga}.

For benchmarking these functionals, we used \gls{sa-casscf} wave functions published in our previous study \cite{KingLargescale2022}, where the approximate pair coefficient scheme \cite{KingRanked2021} was applied to select the active space. Specifically, we used the ``Aug(12,12)'' wave functions, which are constructed with the aug-cc-pVTZ basis set \cite{DunningGaussian1989, KendallElectron1992} and where, as explained in more detail in the previous 
paper\cite{KingLargescale2022}, ``(12,12)'' means that the an active space has fewer configurations than a (12e, 12o) active space.

The accurate results used for benchmarking are the QUEST database that contains \gls{tbe} energies of 542 vertical excitations of main-group molecules containing up to 10 non-hydrogen atoms\cite{LoosMountaineering2018, LoosMountaineering2020,LoosMountaineering2020a, LoosQuest2020, VerilQUESTDB2021}. The \gls{nevpt2} results are from strongly contracted \gls{nevpt2} \cite{AngeliIntroduction2001}.

\section{Results and Discussion}

\subsection{Numerical validation of nuclear gradients}

We first verify our implementation of \gls{mcpdft} analytic nuclear gradients with a \gls{ss-casscf} and \gls{sa-casscf} reference wave functions using a hybrid \gls{ga} (tPBE0) and a hybrid \gls{mga} (MC23) functional by comparing the analytic and numerical nuclear gradients for the diatomic system \ce{LiH} as functions of the \ce{Li-H} bond distance. We calculated the numerical nuclear gradients using the central-difference method with a step size of $\delta$.
\begin{equation}
    \mathrm{Num}\pqty{\delta} = \frac{\energy^\mathrm{PDFT}\pqty{\lambda + \delta} - \energy^\mathrm{PDFT}\pqty{\lambda - \delta}}{2\delta}
\end{equation}
This method has an error of $\order{\delta^2}$; and since it depends explicitly on the step-size $\delta$, we extrapolate to the $\delta \to 0$ limit by taking a sequence of numerical gradients with different $\delta$ values and performing a linear regression on the plot of numerical gradient versus $\delta^2$. The $y$-intercept from this regression is taken to be the best estimate of the numerical gradient. 

Because both the analytic and numerical gradients depend on a wave function that can only be converged to a finite precision, there is inherent error in both values. Regardless, consistently with our prior work \cite{SandAnalytic2018, ScottAnalytic2020, baoAnalytic2022, HennefarthAnalytic2024}, we refer to the numerical value as the reference for the rest of the manuscript. 

\begin{table}
    \footnotesize
    \centering
    \caption{\label{tab:num-err-stats} \Glsxtrfull{mse}, \glsxtrfull{mue}, and 
    \glsxtrfull{rmse} of the analytic gradients (in \unit{\hartree\per\bohr}) relative to the numerical gradients for \ce{LiH}}
    \begin{tabular*}{\columnwidth}{@{\extracolsep{\fill}} l S[table-format=-1.1e-1] S[table-format=-1.1e-1] S[table-format=-1.1e-1] S[table-format=-1.1e-1]}
        \toprule\toprule
        & \multicolumn{2}{c}{\ce{State-Specific}} & \multicolumn{2}{c}{State-Averaged}                              \\\cmidrule{2-3}\cmidrule{4-5}
         Functional    & {tPBE0}  & {MC23}   & {tPBE0}  & {MC23}   \\
         \midrule
         MSE           &  7.5e-07 & -2.2e-06 &  2.6e-06 &  2.8e-06 \\
         MUE           &  4.1e-06 &  6.1e-06 &  6.6e-06 &  6.7e-06 \\
         RMSE          &  7.4e-06 &  1.0e-05 &  9.4e-06 &  1.0e-05 \\
         \bottomrule\bottomrule
    \end{tabular*}
\end{table}

\Cref{tab:num-err-stats} summarizes the statistical agreement between the analytic and numerical gradients for \ce{LiH} using SS-\gls{mcpdft} or SA-\gls{mcpdft}. SS-\gls{mcpdft} has a \gls{mue} of \qty{4.1e-6}{\hartree\per\bohr} for the tPBE0 functional and \qty{6.1e-6}{\hartree\per\bohr} for the  MC23 functional. SA-\gls{mcpdft} has MUEs of \qty{6.6e-6}{\hartree\per\bohr} and \qty{6.7e-6}{\hartree\per\bohr} for the tPBE0 and MC23 functionals respectively. For all cases, the \gls{mue} is on the order of \qty{1e-6}{\hartree\per\bohr}, which is consistent with the agreement between the analytic and numerical nuclear gradients from our prior work \cite{ScottAnalytic2020, baoAnalytic2022, HennefarthAnalytic2024}. See section S1 and fig. S1-S4 for a detailed comparison of the analytical and nuclear gradients.

Overall, we find good agreement between the analytic and numerical gradients for \ce{LiH} using either a translated hybrid \gls{ga} or hybrid \gls{mga} on-top functional with either SS-\gls{mcpdft} or SA-\gls{mcpdft}.

\subsection{The \textit{s-trans}-butadiene molecule}

The \textit{s-trans}-butadiene molecule is known to have significant multireference character in both its ground and excited $\prescript{1}{}{\mathrm{A}}_\mathrm{g}$ states \cite{ShuDoubly2017}, and we use its ground- and excited-state geometries as our first optimization tests. \Cref{tab:mga-butadiene-geometry} presents the optimized and reference values for the key internal coordinates of the 1 $\prescript{1}{}{\mathrm{A}}_\mathrm{g}$ and 2 $\prescript{1}{}{\mathrm{A}}_\mathrm{g}$ states of \textit{s-trans}-butadiene computed with various on-top functionals. We use a (4e, 4o) active space consisting of $2 \pi$ and $2 \pi^*$ orbitals. For the ground-state geometry reference, we take the experimental values from \citet{Haugenmolecular1966}, and for the excited-state reference, we take our prior SA(2)-CASPT2(4,4) calculation \cite{ScottAnalytic2020}. We also include ground-state results from \gls{cc3} theory \cite{LoosReference2019}.

\begin{table*}
  \centering
  \caption{\label{tab:mga-butadiene-geometry} Selected optimized internal coordinates (in \unit{\angstrom} and degrees) for the ground and excited $\prescript{1}{}{\mathrm{A}}_\mathrm{g}$ states of \textit{s-trans}-butadiene}
	\begin{tabular*}{\textwidth}{!{\extracolsep\fill} l l S[table-format=1.3] S[table-format=1.3] S[table-format=3.1] S[table-format=1.3] S[table-format=1.3] S[table-format=3.1]}
		\toprule\toprule
        & & \multicolumn{3}{c}{1 \textsuperscript{1}A\textsubscript{g}} & \multicolumn{3}{c}{2 \textsuperscript{1}A\textsubscript{g}} \\ \cmidrule{3-5} \cmidrule{6-8}
		Method & Basis Set & $r_{\ce{C=C}}$ & $r_{\ce{CC}}$ & $\theta_{\ce{CCC}}$ & $r_{\ce{C=C}}$ & $r_{\ce{CC}}$ & $\theta_{\ce{CCC}}$ \\
		\midrule
        Pair-density functional theory\\
        SA(2)-tPBE(4,4)     & jul-cc-pVTZ & 1.336 & 1.459 & 124.1 & 1.497 & 1.398 & 124.1 \\
        SA(2)-tPBE0(4,4)    & jul-cc-pVTZ & 1.338 & 1.458 & 124.2 & 1.497 & 1.402 & 123.9 \\
        SA(2)-MC23(4,4)     & jul-cc-pVTZ & 1.335 & 1.451 & 123.8 & 1.491 & 1.397 & 123.3 \\
        L(2)-tPBE(4,4) \cite{HennefarthAnalytic2024} & jul-cc-pVTZ & 1.335 & 1.460 & 124.1 & 1.496 & 1.399 & 124.1 \\
        \\
        Comparison results\\
        SA(2)-CASSCF(4,4)   & jul-cc-pVTZ & 1.345 & 1.456 & 124.3 & 1.497 & 1.413 & 123.2 \\
        SA(2)-CASPT2(4,4) \cite{ScottAnalytic2020}   & aug-cc-pVTZ & 1.342 & 1.454 & 123.6 & 1.488 & 1.394 & 122.1\\
        \glsxtrshort{cc3} \cite{LoosReference2019}   & aug-cc-pVTZ & 1.340 & 1.453 & 123.9 \\
		exp. \cite{Haugenmolecular1966}  & & 1.343 & 1.467 & 122.8 \\
		\bottomrule\bottomrule
	\end{tabular*}
\end{table*}

\Cref{tab:mga-butadiene-geometry} shows that the internal coordinates of both the ground- and excited-state geometries of \textit{s-trans}-butadiene show minimal variation when switching between the tPBE and tPBE0 on-top functionals. In most cases, as might be expected, the tPBE0 results fall between those obtained with tPBE and \gls{sa-casscf}. For the excited-state geometry, the MC23 functional yields coordinates closest to the \gls{caspt2} reference among all the PDFT methods, which is encouraging because MC23 was optimized solely for ground-state properties and a few spin splittings. It is satisfying that \textit{all} on-top functionals produce similar geometries, with no significant deviations observed for either state.

\Cref{tab:butadiene-energy} compares the vertical and adiabatic excitation energies computed with various on-top functionals to various other high-level electronic-structure results, including \glspl{tbe}. There are two very-high-quality \glspl{tbe} for the vertical excitation energy of \textit{s-trans}-butadiene in the literature. \citet{WatsonExcited2012} estimate the vertical excitation energy to be \qty{6.39}{\eV} using extrapolated equation of motion coupled cluster calculations and \citet{LoosReference2019} estimate the vertical excitation energy to be \qty{6.50}{\eV} using extrapolated complete-basis-set \gls{fci} calculations. They are both designated as \gls{tbe} in \cref{tab:butadiene-energy} since it is not certain which is more accurate. Both \gls{ms-caspt2} (at the experimental geometry)\cite{ShuDoubly2017} and \gls{caspt2} \cite{ScottAnalytic2020} perform similarly for the vertical excitation energy, differing by only \qty{0.01}{\eV}. Similarly, L-tPBE and tPBE differ by only \qty{0.01}{\eV}, showing that state interaction between the 1 $\prescript{1}{}{\mathrm{A}}_\mathrm{g}$ and 2 $\prescript{1}{}{\mathrm{A}}_\mathrm{g}$ states is not important for this excitation (state interaction between closely spaced excited states is usually more important than state interaction between an excited state and the ground state). This also illustrates how the multistate L-PDFT does not degrade the accuracy of single-state MC-PDFT in a case where the states coupled do not interact strongly. The table shows that tPBE0 performs the best of all of the PDFT methods at predicting the vertical excitation energy. 

For the adiabatic excitation energy, we take our prior \gls{caspt2} results as our reference \cite{ScottAnalytic2020}. MC23 performs almost identically to tPBE for both the vertical and adiabatic excitation. Similarly, tPBE0 gives an adiabatic excitation energy identical to that of \gls{caspt2}, making it the best performing on-top functional for predicting the \textit{s-trans}-butadiene excitation energies.

\begin{table}
  \footnotesize
  \centering
  \caption{\label{tab:butadiene-energy} Adiabatic and vertical excitations energies in \unit{\electronvolt} (not including vibrational energy) for the $1\prescript{1}{}{\mathrm{A}}_\mathrm{g} \to 2\prescript{1}{}{\mathrm{A}}_\mathrm{g}$ excitation of \textit{s-trans}-butadiene}
  \begin{threeparttable}
    \begin{tabular*}{\columnwidth}{!{\extracolsep\fill}l l S[table-format=1.2] S[table-format=1.2]}
      \toprule\toprule
      Method & Basis Set & {Vertical} & {Adiabatic} \\
      \midrule
      Pair-density functional theory\\
      SA(2)-tPBE(4,4)     & jul-cc-pVTZ & 6.91 & 5.77 \\
      SA(2)-tPBE0(4,4)    & jul-cc-pVTZ & 6.82 & 5.68 \\
      SA(2)-MC23(4,4)     & jul-cc-pVTZ & 6.91 & 5.78 \\
      L(2)-tPBE(4,4) \cite{HennefarthAnalytic2024} & jul-cc-pVTZ & 6.92 & 5.78 \\
      \\
      Comparison results \\
      SA(2)-CASSCF(4,4)  & jul-cc-pVTZ & 6.56 & 5.42\\
      SA(2)-CASPT2(4,4) \cite{ScottAnalytic2020}  & aug-cc-pVTZ & 6.68 & 5.68\tnote{b} \\
      MS(2)-CASPT2(4,4)\tnote{a,} \cite{ShuDoubly2017} & 6-31G** & 6.69 & \\
      \glsxtrshort{ccsd}\tnote{a,} \cite{ShuDoubly2017} & 6-31G** & 7.69 & \\
      \glsxtrshort{cc3} \cite{LoosReference2019} & aug-cc-pVTZ & 6.67 & \\
      \citet{WatsonExcited2012}  & & 6.39\tnote{b} & \\
      \citet{LoosReference2019} & & 6.50\tnote{b} & \\
      \bottomrule\bottomrule
    \end{tabular*}
    \begin{tablenotes}
      \footnotesize
      \item[a] Excitation energy calculated at experimental equilibrium geometry
      \item[b] \Glsentrylong{tbe}
    \end{tablenotes}
  \end{threeparttable}
\end{table}

\subsection{Benzophenone}

\begin{figure}
    \centering
    \includegraphics[width=\columnwidth]{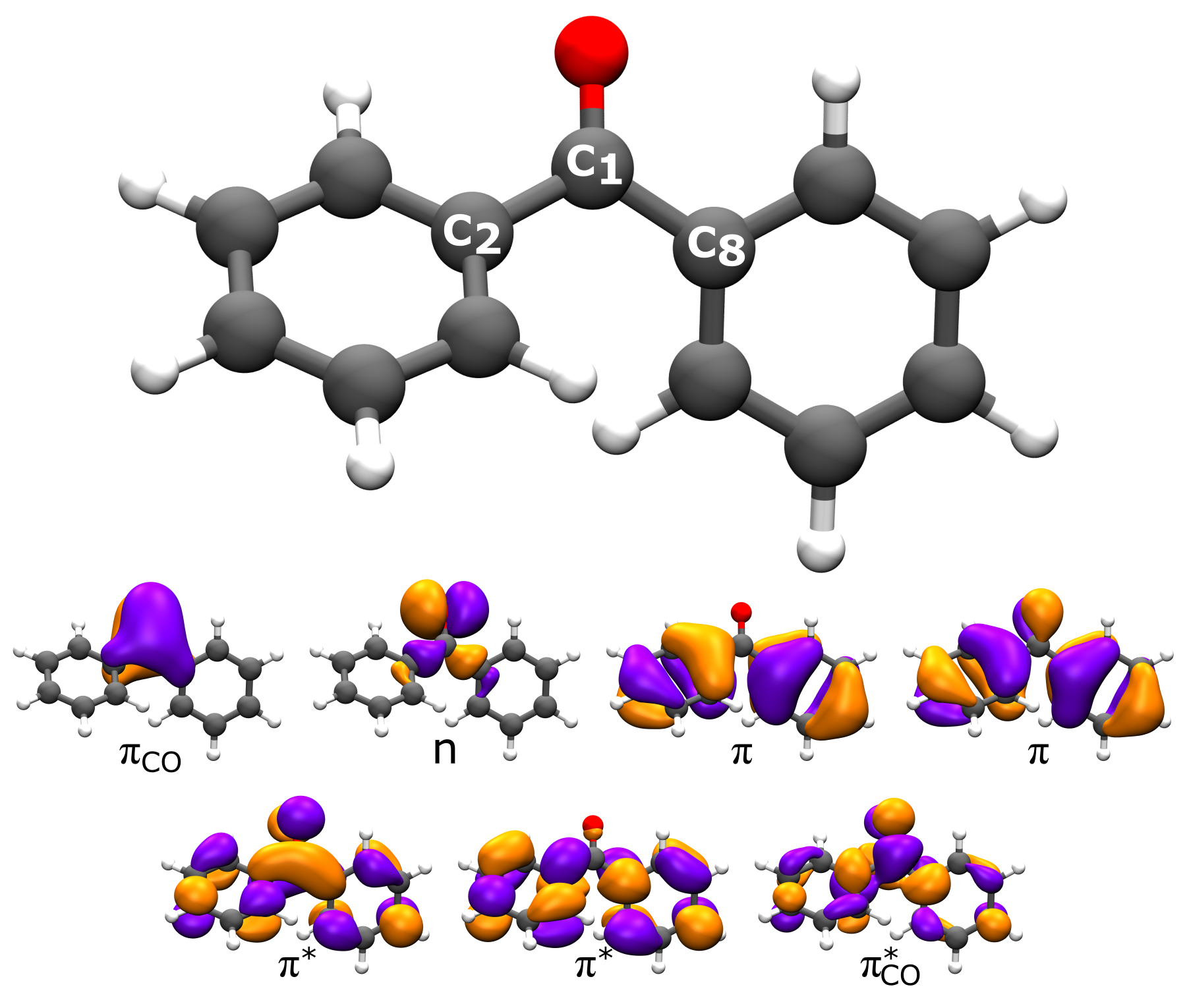}
    \caption{\label{fig:benzophenone-structure} Ground-state geometry of benzophenone optimized with SA(3)-tPBE(8,7), and the orbitals in the (8e,7o) active space.}
\end{figure}

Next we discuss the ground- and first-singlet excited-state geometry of benzophenone (\cref{fig:benzophenone-structure}). Benzophenone is an important triplet photosensitizer and is a prototype for the photochemistry of more complex aromatic ketones. It is known that transitions from the $S_1$ state to the triplet manifold proceed with near-unit quantum yield in single crystals, solutions, and isolated matrices \cite{AloiseBenzophenone2008, KatohTriplet1997, OhmoriWhy1988}. Benzophenone has been studied extensively both experimentally and computationally. 

It is known that the ground- and first-singlet excited-state geometries have $C_2$ symmetry. There is a balance between conjugation and steric effects between the two phenyl rings. The conjugation favors coplanarity whereas the steric effects between the phenyl hydrogen's favors breaking the coplanarity. The $S_0$ to $S_1$ geometry change is characterized by an elongation of the \ce{C=O} bond and a reduction of the angle between the phenyl rings (which we denote as $\tau$) \cite{HoffmannGround1970, SergentuComputational2014}. 

\begin{table*}
  \centering
  \begin{threeparttable}
    \caption{\label{tab:mga-benzophenone-geometry} Selected optimized internal coordinates for the ground and excited $\prescript{1}{}{\mathrm{A}}$ states of benzophenone\tnote{a}} 
	\begin{tabular*}{\textwidth}{!{\extracolsep\fill} l S[table-format=1.3] S[table-format=1.3] S[table-format=3.1] S[table-format=2.1] S[table-format=1.3] S[table-format=1.3] S[table-format=3.1] S[table-format=2.1]}
		\toprule\toprule
        & \multicolumn{4}{c}{1 \textsuperscript{1}A} & \multicolumn{4}{c}{2 \textsuperscript{1}A} \\ \cmidrule{2-5} \cmidrule{6-9}
		Method  & $r_{\ce{C=O}}$ & $r_{\ce{C1C2}}$ & $\theta_{\ce{C2C1C8}}$ & $\tau$ & $r_{\ce{C=O}}$ & $r_{\ce{C1C2}}$ & $\theta_{\ce{C2C1C8}}$ & $\tau$ \\
		\midrule
        SA(3)-CASSCF & 1.213 & 1.501 & 121.2 & 58.5 & 1.369 & 1.445 & 130.9 & 38.8 \\
        SA(3)-tPBE   & 1.218 & 1.514 & 119.7 & 55.6 & 1.353 & 1.444 & 129.9 & 37.8 \\
        SA(3)-tPBE0  & 1.217 & 1.510 & 120.1 & 56.4 & 1.358 & 1.444 & 130.2 & 38.0 \\
        SA(3)-MC23   & 1.216 & 1.502 & 119.9 & 54.5 & 1.355 & 1.437 & 130.1 & 37.4 \\
        XMS(3)-CASPT2 \cite{VlaisavljevichNuclear2016}  & 1.231 & 1.500 & 119.1 & 52.9 & 1.350 & 1.444 & 128.6 & 39.4 \\
		exp. \cite{FleischerCrystal1968} & 1.231 & 1.496 & 121.7 & 55.5 \\
		\bottomrule\bottomrule
	\end{tabular*}
    \begin{tablenotes}
        \footnotesize
        \item[a]  Atoms are numbered according to \cref{fig:benzophenone-structure}. All bond lengths are in \unit{\angstrom} and all angles are in degrees. $\tau$ is defined as the angle between the two planes of the phenyl rings. All calculations in this table use an (8e,7o) active space and the cc-pVDZ basis set.
    \end{tablenotes}
    \end{threeparttable}
\end{table*}

We use an (8e, 7o) active space comprising an oxygen nonbonding orbital, the \ce{C=O} $\pi$ and $\pi^*$ orbitals, and $2 \pi$ and $2 \pi^*$ orbitals on each of the phenyl groups (\cref{fig:benzophenone-structure}). The first singlet excitation is of $n \to \pi^*$ character with both the ground- and excited-state wave functions having $\prescript{1}{}{\mathrm{A}}$ symmetry. \Cref{tab:mga-benzophenone-geometry} summarizes selected optimized internal coordinates for these two $\prescript{1}{}{\mathrm{A}}$ states. For the ground-state geometry we take the experimental structure from \citet{FleischerCrystal1968} as our reference, and for the excited-state geometry we take the \gls{xms-caspt2} structure \cite{VlaisavljevichNuclear2016} as our reference. 

For the ground state, \cref{tab:mga-benzophenone-geometry} shows that \gls{sa-casscf} and all of the PDFT methods underestimate the \ce{C=O} bond length whereas \gls{xms-caspt2} agrees with the experimental value of \qty{1.231}{\angstrom}. For the \ce{C1-C2} bond, MC23 and \gls{xms-caspt2} get bond length close to the experimental one. For the \ce{C2-C1-C8} bond angle, \gls{sa-casscf} gets the closest to the experimental quantity followed by tPBE0. For the angle between the two phenyl rings ($\tau$), MC23 gets the closest to the experimental \qty{55.5}{\degree}. 

For the excited-state geometry, \cref{tab:mga-benzophenone-geometry} shows that \gls{sa-casscf} significantly overestimates the \ce{C=O} bond length, and all three PDFT methods reduce the error by more than a factor of two, with tPBE getting closest to the \gls{xms-caspt2} bond length. For the \ce{C1-C2} bond length, both tPBE and tPBE0 agree within \qty{0.001}{\angstrom} with \gls{xms-caspt2}, whereas MC23 predicts a slightly shorter bond length by \qty{0.007}{\angstrom}. The \gls{sa-casscf} method, tPBE, and tPBE0 predict that the angle between the phenyl rings is smaller in the excited-state than in the ground state by \qtyrange{18}{20}{\degree}, MC23 predicts it is smaller by 1\qty{17}{\degree}, and \gls{xms-caspt2} predicts it is smaller by \qty{13.5}{\degree}. In summary, there are no large deviations between any of the on-top functionals in predicting the ground- and excited-state structures of benzophenone with all methods in general agreement with both the experimental value and \gls{xms-caspt2}. 

\Cref{tab:mga-benzophenone-energy} summarizes the adiabatic and vertical excitation energies of benzophenone as calculated with various on-top functionals. An experimental quantity from \citet{ItohEmission1985} suggests that the vertical excitation energy is \qty{3.61}{\eV}, which \gls{caspt2} with a larger (12e, 11o) and (16e, 15o) active space (both calculated at the \gls{sa-casscf} optimized geometry) gets to within \qty{0.05}{\eV} of \cite{SergentuComputational2014}. Both tPBE and tPBE0 predict a vertical excitation of \qty{3.78}{\eV} with an error of \qty{0.17}{\eV} relative to the experimental value. A prior calculation using \gls{ccsd} predicts an even higher excitation with an error of \qty{0.39}{\eV}, worse than \gls{sa-casscf}. For the adiabatic excitation, we take the \gls{xms-caspt2} as our reference value. All PDFT calculations lie between the \gls{xms-caspt2} and \gls{caspt2} results. tPBE gets the closest to the \gls{xms-caspt2} result, followed by MC23. The performance of tPBE0 is hampered due to \gls{sa-casscf} predicting a lower adiabatic excitation energy, with a deviation of \qty{0.32}{\eV} relative to the \gls{xms-caspt2} value.

\begin{table}
  \centering
  \footnotesize
  \caption{\label{tab:mga-benzophenone-energy} Adiabatic and vertical excitation energies in \unit{\electronvolt} (not including vibrational zero-point energy) for the $1\prescript{1}{}{\mathrm{A}} \to 2\prescript{1}{}{\mathrm{A}}$ excitation of benzophenone}
  \begin{threeparttable}
    \begin{tabular*}{\columnwidth}{!{\extracolsep\fill}l l S[table-format=1.2] S[table-format=1.2]}
      \toprule\toprule
      Method & Basis set & {Vertical} & {Adiabatic} \\
      \midrule
      SA(3)-CASSCF(8,7) & cc-pVDZ & 3.83 & 3.04 \\
      SA(3)-tPBE(8,7)   & cc-pVDZ & 3.78 & 3.25 \\
      SA(3)-tPBE0(8,7)  & cc-pVDZ & 3.78 & 3.19 \\
      SA(3)-MC23(8,7)   & cc-pVDZ & 3.82 & 3.22 \\
      SA(4)-CASPT2(12,11)\tnote{a,} \cite{SergentuComputational2014} & ANO-L-VDZP & 3.64 & 3.15 \\
      SA(4)-CASPT2(16,15)\tnote{a,} \cite{SergentuComputational2014} & ANO-L-VDZP & 3.66 & \\
      XMS(3)-CASPT2(8,7) \cite{VlaisavljevichNuclear2016} & cc-pVDZ & & 3.36\\
      \glsxtrshort{ccsd}\tnote{a,} \cite{SergentuComputational2014} & cc-pVDZ & 4.00 & \\
      exp. \cite{ItohEmission1985}   & & 3.61 & \\
      \bottomrule\bottomrule
    \end{tabular*}
    \begin{tablenotes}
        \footnotesize
        \item[a] Excitation energy calculated at SA(4)-CASSCF(12/11)/ANO-L-VDZP optimized geometries.
    \end{tablenotes}
  \end{threeparttable}
\end{table}

\subsection{Performance of \glsentrylong{mga} functionals in predicting vertical excitation energies in the QUEST database}

\begin{figure}
    \centering
    \includegraphics[width=\columnwidth]{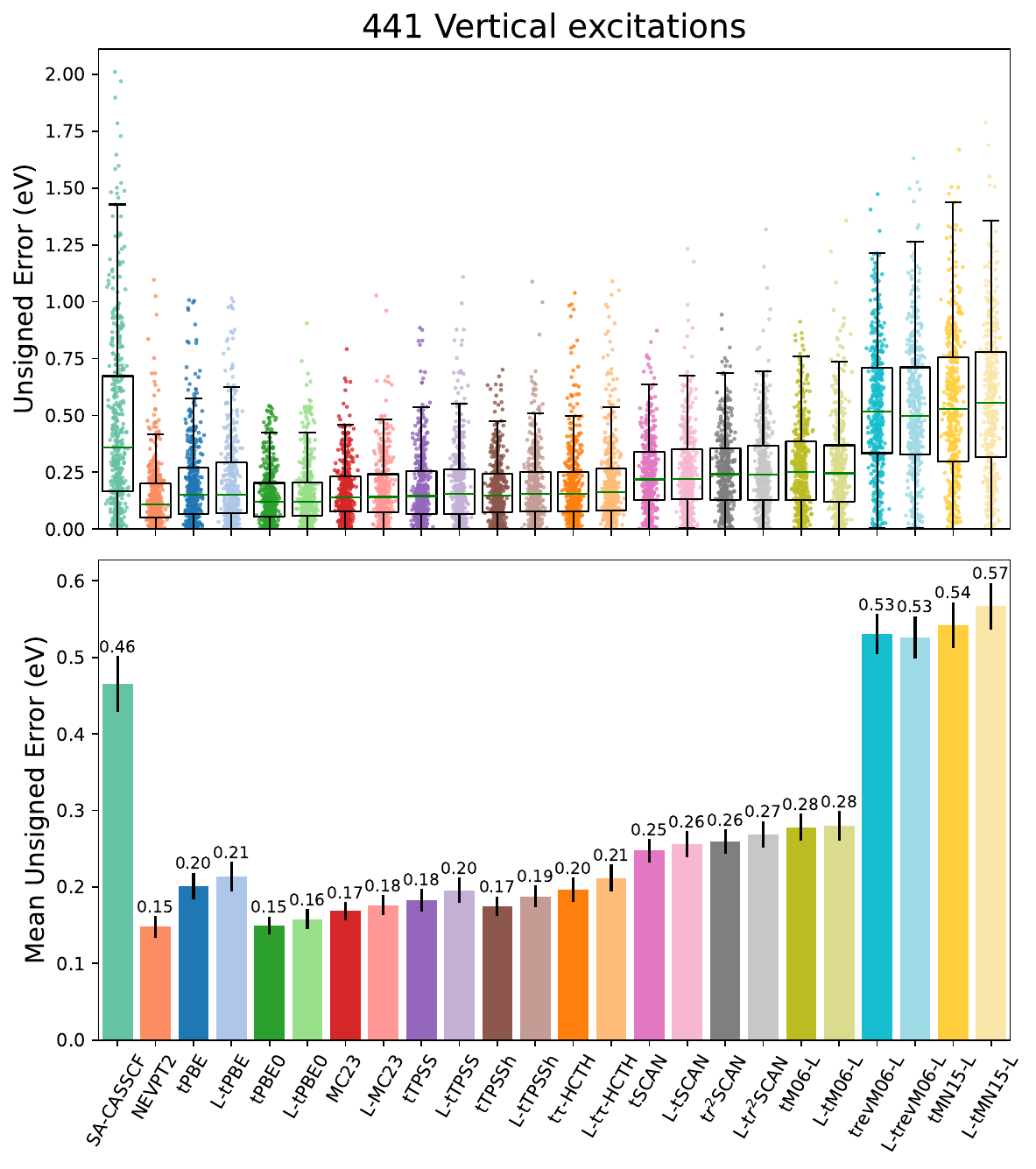}
    \caption{\label{fig:questdb-tpbe0-cutoff} Error statistics across the 441 excitations for \glsxtrshort{sa-casscf}, \glsxtrshort{nevpt2}, and 11 functionals for \glsxtrshort{mcpdft} and \glsxtrshort{lpdft}. Top: Distribution of vertical excitation \gls{ue} and box and whisker plots.  Each dot represents a single \gls{ue} for the given method. The box represents the inter-quartile range and the horizontal green lines within each box indicates the median value. The whiskers extend out $1.5$ times the inter-quartile range. Bottom: Bar plots of the \gls{mue} for each method on the 441 excitations with the black vertical lines indicating the \qty{95}{\percent} confidence intervals. \glsxtrshort{sa-casscf}, tPBE, tPBE0, and \glsxtrshort{nevpt2} data are taken from \citet{KingLargescale2022}; and L-tPBE and L-tPBE0 data are taken from \citet{HennefarthLinearized2023a}}
\end{figure}

Here, we study \gls{mga} functionals with \gls{mcpdft} and \gls{lpdft} for the prediction of  vertical excitation energies in a dataset of 441 excitations for which accurate results are available in the QUEST database,\cite{LoosReference2019, LoosMountaineering2018, LoosMountaineering2020, LoosMountaineering2020a, VerilQUESTDB2021, LoosQuest2020} and reference \gls{sa-casscf} wave functions are available from a previous study \cite{KingLargescale2022}. Excitations for which a tPBE0 excitation energy calculated with these available \gls{sa-casscf} reference wave functions has an unsigned deviation from the reference value greater than \qty{0.55}{\eV} have been removed from the original set of 520 excitations (giving the subset of 441 excitations) because that may be an indication that the active spaces of those reference wave functions are inadequate. In this way, we are testing only with \gls{sa-casscf} wave functions that should be adequate. The distribution of \gls{ue} and \gls{mue} for the 441 retained excitations are shown in \cref{fig:questdb-tpbe0-cutoff}. A plot for the entire 540 excitations is shown in fig. S5 in the supporting information.

We study one translated \gls{ga} (tPBE), one hybrid translated \gls{ga} (tPBE0), seven translated \glspl{mga} (tTPSS, t$\tau$-HCTH, tSCAN, tr$^2$SCAN, tM06-L, trevM06-L, and tMN15-L) and two hybrid \glspl{mga} (tTPSSh and MC23).
\Cref{fig:questdb-tpbe0-cutoff} shows that tPBE0 (with an \gls{mue} of \qty{0.15}{\eV}) exhibits the best overall performance, matching that of \gls{nevpt2} (\gls{mue} \qty{0.15}{\eV}) for the subset selected by the tPBE0 error threshold, and fig. S5 shows that it also performs best for the entire data set without filtering, although in that case the \gls{mue} of \qty{0.44}{\eV} is much higher. Among the tested translated \gls{mga} and hybrid translated \gls{mga} functionals, MC23 shows the best performance (\gls{mue} \qty{0.17}{\eV}), followed by tTPSS (\qty{0.18}{\eV}), tTPSSh (\qty{0.18}{\eV} eV), t$\tau$-HCTH (\qty{0.20}{\eV}), tSCAN (\qty{0.25}{\eV}), tr$^2$SCAN (\qty{0.26}{\eV}), tM06-L (\qty{0.28}{\eV}), trevM06-L (\qty{0.53}{\eV}), and tMN15-L (\qty{0.54}{\eV}).

Comparisons between \gls{lpdft} and \gls{mcpdft} results for each functional reveal a strong correspondence, with a difference in \glspl{mue} generally within \qty{0.02}{\eV}. While \cref{fig:questdb-tpbe0-cutoff} might suggest that \gls{lpdft} performs consistently worse (by 0.01 eV) than \gls{mcpdft} across all functionals, this is a consequence of the thresholding based on the tPBE0 error. When thresholding is performed using the L-tPBE0 error, the trend is reversed: \gls{lpdft} shows equal or improved accuracy for most functionals (fig. S6). These results indicate that \gls{lpdft} successfully reproduces \gls{mcpdft} performance within the scope of this benchmark and can be routinely applied to study excitations. The findings are consistent with our previous study, where tPBE and tPBE0 were tested \cite{HennefarthLinearized2023a}.

\subsection{Comparing \glsxtrshort{mcpdft} to \glsxtrshort{ks-dft}}

\begin{figure}
    \centering
    \includegraphics[width=\columnwidth]{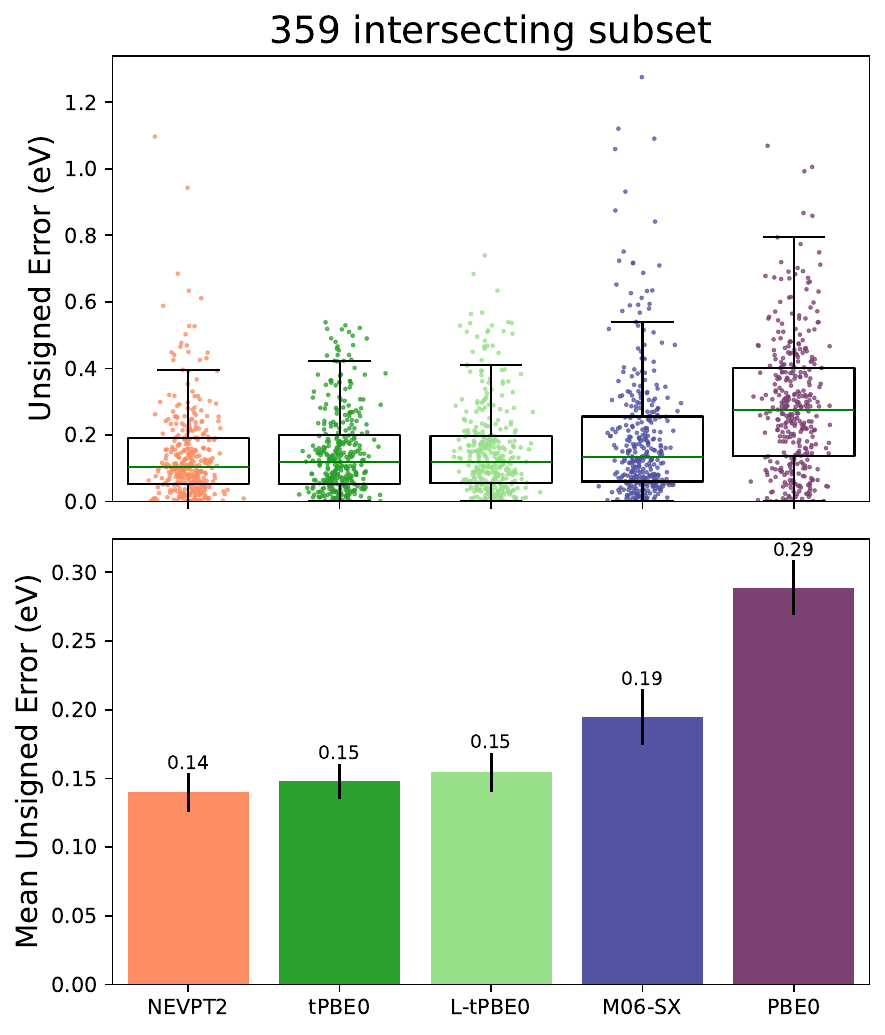}
    \caption{\label{fig:questdb-ksdft-comparison} Error statistics comparison between \citet{LiangRevisiting2022} and our study across the intersecting subset of 359 excitations. Results are shown for \gls{nevpt2}, tPBE0, and L-tPBE0 from our study as well as the M06-SX and PBE0 \gls{ks-dft} functionals.}
\end{figure}

We also compare the performance of \gls{mcpdft} to the performance of \gls{tddft} as studied by \citet{LiangRevisiting2022}, who tested more than 40 \gls{ks-dft} functionals on a 463-excitation subset of QUESTDB. Due to differing selection criteria, the subsets used in our study and theirs are not identical. Therefore, we focus on the intersection of the two data sets, which comprises 359 excitations (listed in supporting information SIII); this enables a direct comparison between the two benchmark studies. The distributions of \glspl{ue} and \glspl{mue} for each study's best-performing methods are shown in \cref{fig:questdb-ksdft-comparison}, while the \glspl{mue} of all methods tested in our study and in \citet{LiangRevisiting2022} are shown in table S1 and S2, respectively. The best performing \gls{ks-dft} functional is M06-SX \cite{WangM062020} with an \gls{mue} of \qty{0.20}{\eV}, while the best on-top functional in our study is tPBE0, which achieves an \gls{mue} of \qty{0.15}{\eV}. This is three-quarters of the error of M06-SX and about half of the error of tPBE0's \gls{ks-dft} counterpart PBE0 (\gls{mue} \qty{0.29}{\eV} for the intersecting subset), suggesting that \gls{mcpdft} performance surpasses \gls{tddft} when \gls{mcpdft} uses a well-chosen active space. Moreover, the performance of tPBE0 aligns closely with that of \gls{nevpt2} (\gls{mue} \qty{0.14}{\eV} on the intersecting subset), suggesting that equally high accuracy can be obtained at lower cost by using \gls{mcpdft}.

\section{Conclusion}

We derived the analytic nuclear gradients of SS-\gls{mcpdft} and SA-\gls{mcpdft} (with and without density fitting) for \gls{mga} and hybrid \gls{mga} on-top functionals. We verified our implementation by showing the analytic and numerical gradients for \ce{LiH} agree to better than \qty{1e-5}{\hartree\per\bohr}. Then we showed the utility of such nuclear gradients by performing geometry optimizations for the ground state and first excited singlet state of \textit{s-trans}-butadiene and benzophenone. Because the derivation of the \gls{mcpdft} nuclear gradients for \gls{mga} on-top functionals produced the one-electron and two-electron on-top potentials for a \gls{mga}, it also enabled us to calculate \gls{lpdft} energies with \gls{mga} on-top functionals. We used this capability to benchmark seven \gls{mga} on-top functionals and two hybrid \gls{mga} on-top functionals on the QUEST dataset of vertical excitation energies, and we found that the MC23 \gls{mga} functional performed the best of these nine functionals and is on par with the hybrid \gls{ga} functional tPBE0.

A next step is to enable \gls{lpdft} analytic nuclear gradients with a \gls{mga} on-top functional. This requires taking another derivative of the non-linear translation scheme. 

\section*{Supplementary Material}
Detailed analysis of numerical versus analytical gradients. Vertical excitation energy for each method and system within the QUESTDB (csv file). Plots of vertical excitation benchmarking for full QUEST dataset as well as L-tPBE0 thresholded systems and 359-excitation overlap subset with prior \gls{tddft} calculations; the 359 excitations used to compare \gls{tddft} and \gls{mcpdft}; optimized \textit{s-trans}-butadiene and benzophenol coordinates with total energies.

\begin{acknowledgments}
This work was supported in part by the Air Force Office Scientific Research (grant no. FA9550-20-1-0360). M. R. Hennefarth acknowledges support by the National Science Foundation Graduate Research Fellowship under Grant No. 2140001. We also acknowledge the University of Chicago’s Research Computing Center for their support of this work. Any opinion, findings, and conclusions or recommendations expressed in this material are those of the author(s) and do not necessarily reflect the views of the National Science Foundation.
\end{acknowledgments}

\bibliographystyle{achemso}
\bibliography{ms}

\section*{TOC Graphic}
\includegraphics[width=\columnwidth]{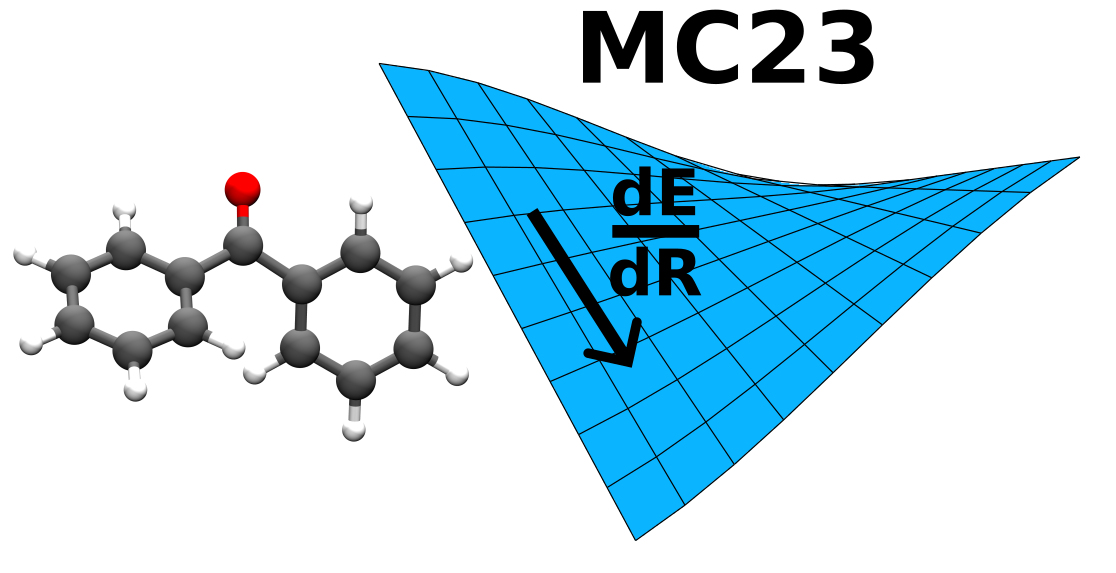}

\end{document}


\title{Supporting Information: MC-PDFT Nuclear Gradients and L-PDFT Energies with Meta and Hybrid Meta On-Top Functionals for Ground- and Excited-State Geometry Optimization and Vertical Excitation Energies}

\author{Matthew R. Hennefarth}
\author{Younghwan Kim}
\author{Bhavnesh Jangid}
\author{Jacob Wardzala}
\author{Matthew R. Hermes}
\affiliation{Department of Chemistry and Chicago Center for Theoretical Chemistry, University of Chicago, Chicago, IL 60637, USA}

\author{Donald G. Truhlar} \email[corresponding author: ]{truhlar@umn.edu}
\affiliation{Department of Chemistry, Chemical Theory Center, and Minnesota Supercomputing Institute, University of Minnesota, Minneapolis, MN 55455-0431, USA}

\author{Laura Gagliardi} \email[corresponding author: ]{lgagliardi@uchicago.edu} 
\affiliation{Department of Chemistry and Chicago Center for Theoretical Chemistry, University of Chicago, Chicago, IL 60637, USA}
\affiliation{Pritzker School of Molecular Engineering, University of Chicago, Chicago, IL 60637, USA}

\date{August 18, 2025}

\maketitle

\tableofcontents

\section{Validation of Analytical Gradients}

We define the \gls{ue} and \gls{ure} for the nuclear gradients as
\begin{equation}
    \mathrm{UE} = \abs{\mathrm{Analytic} - \mathrm{Numerical}}
\end{equation}
\begin{equation}
    \mathrm{URE} = \abs{\frac{\mathrm{Analytic} - \mathrm{Numerical}}{\mathrm{Numerical}}}
\end{equation}

\begin{figure}
    \centering
    \includegraphics[width=0.95\columnwidth]{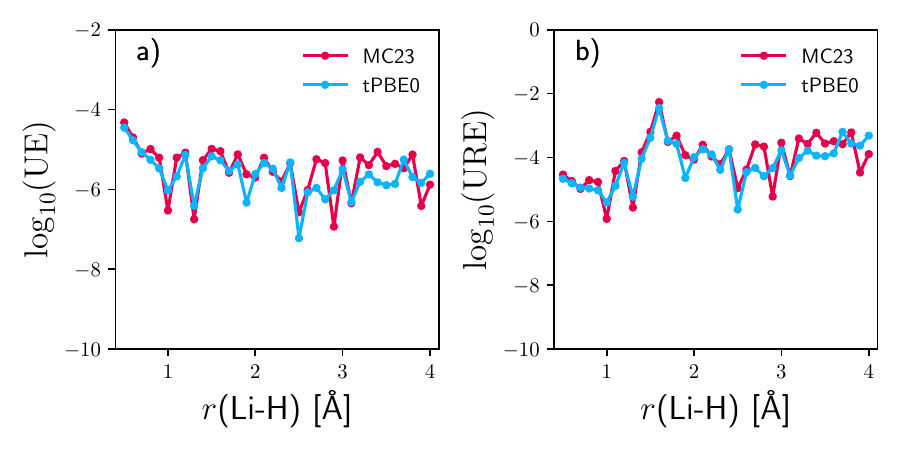}
    \caption{\label{sfig:ss-num-vs-ana-curves} Common $\log$ of the \glsxtrshort{ue} in atomic units (a) and \glsxtrshort{ure} (b) of the analytic gradients relative to the numerical gradients for the ground state of \ce{LiH} using a \glsxtrshort{ss-casscf} reference wave function.}
\end{figure}

\begin{figure}
    \centering
    \includegraphics[width=0.95\columnwidth]{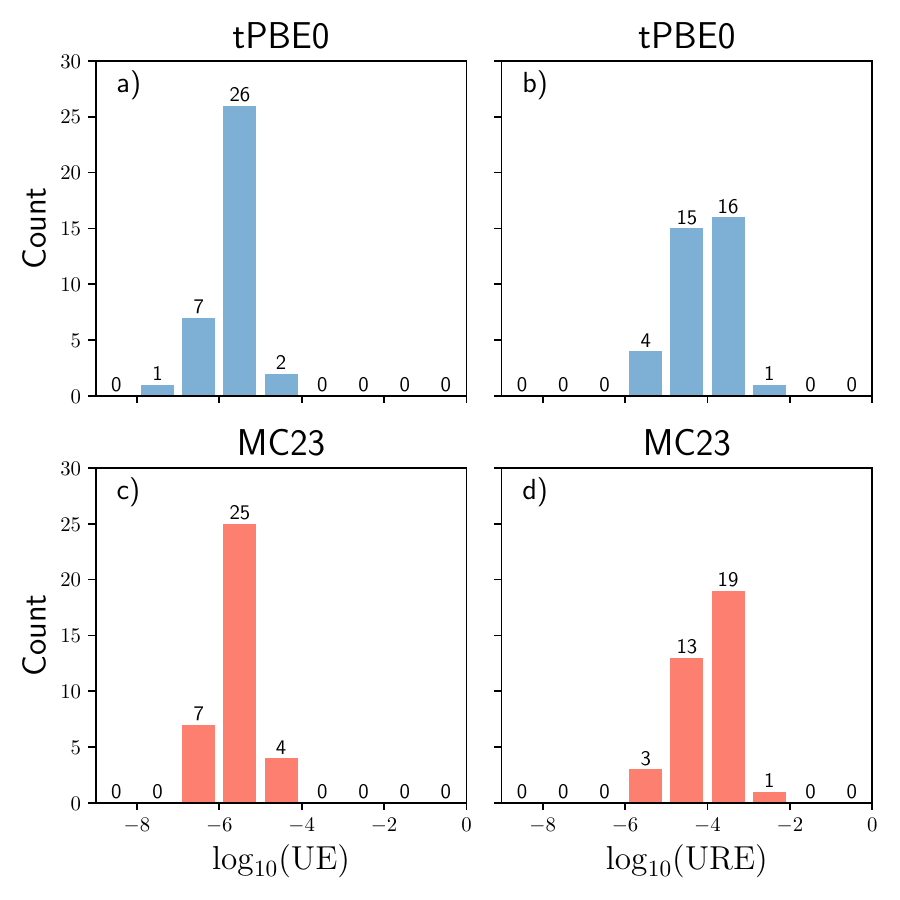}
    \caption{\label{sfig:ss-err-distribution} Distribution of the common $\log$ of the \glsxtrshort{ue} in atomic units (a,c) and \glsxtrshort{ure} (b,d) of the analytic gradients relative to the numerical gradients for the ground state of \ce{LiH} using a \glsxtrshort{ss-casscf} reference wave function. The top row (a,b) uses the tPBE0 function, and the bottom row (c,d) uses the MC23 functional.}
\end{figure}

The common $\log$ of the \glsxtrshort{ue} and \glsxtrshort{ure} for the nuclear gradient of the ground state of \ce{LiH} scanned between \qtyrange{0.5}{4.0}{\angstrom} with a step size of \qty{0.1}{\angstrom} using SS-MC-PDFT with tPBE0 (a hybrid \glsxtrshort{ga} on-top functional) and MC23 (a hybrid \glsxtrshort{mga} on-top functional) is shown in \cref{sfig:ss-num-vs-ana-curves}. This figure shows that the \glsxtrshort{ue} is stable at all bond distances. We show histograms of the errors in \cref{sfig:ss-err-distribution}. This figure shows that \glsxtrshort{ue} is below \qty{1e-4}{\hartree\per\bohr} with the majority being on the order of \qty{1e-6}{\hartree\per\bohr} (\cref{sfig:ss-err-distribution}). The \glsxtrshort{ure} tends to grow as the bond distance increases because the  potential energy curve becomes flatter, but it never goes above \qty{1e-2}{}.

\begin{figure}
    \centering
    \includegraphics[width=0.95\columnwidth]{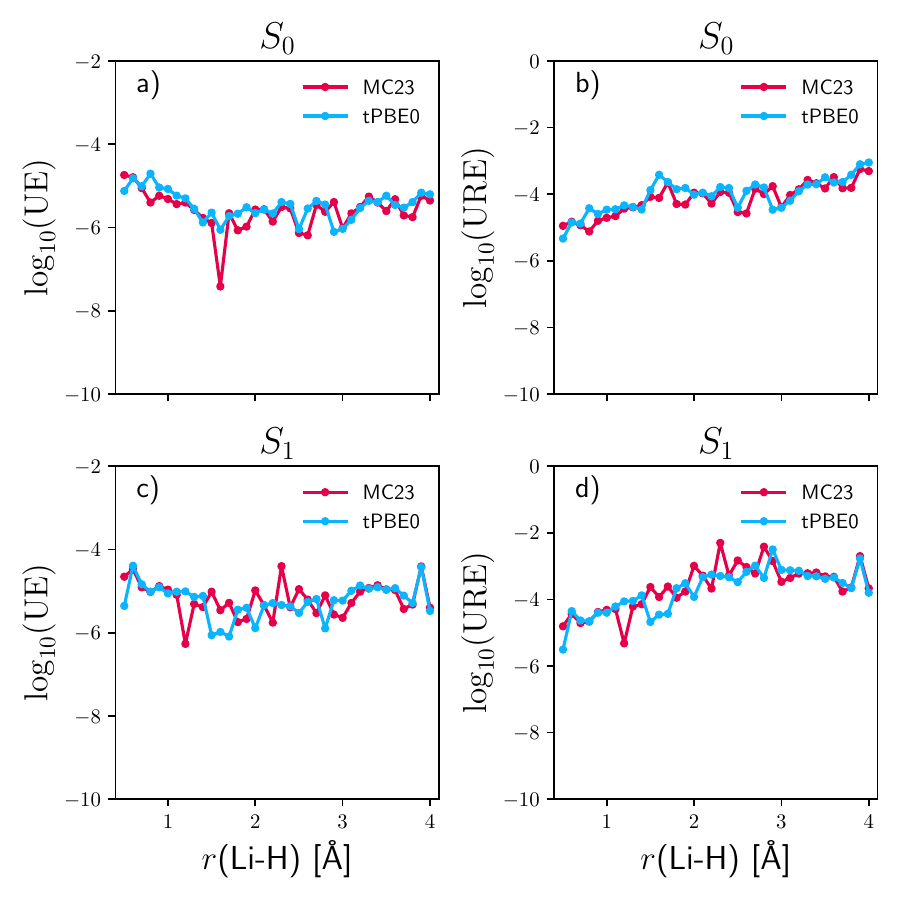}
    \caption{\label{sfig:sa-num-vs-ana-curves} Common $\log$ of the \glsxtrshort{ue} in atomic units (a,c) and \glsxtrshort{ure} (b,d) of the analytic gradients relative to the numerical gradients for the ground (a,b) and first excited (c,d) state of \ce{LiH} using a \glsxtrshort{sa-casscf} reference wave function.}
\end{figure}

\begin{figure}
    \centering
    \includegraphics[width=0.95\columnwidth]{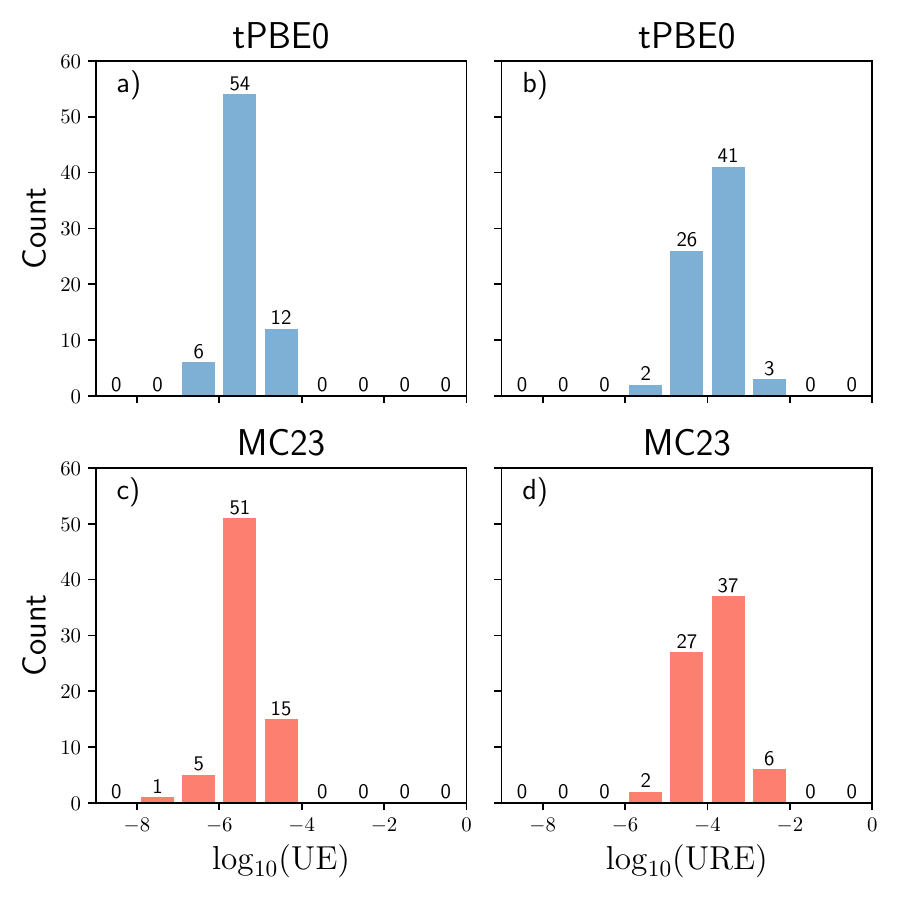}
    \caption{\label{sfig:sa-err-distribution} Distribution of the common $\log$ of the \glsxtrshort{ue} (a,c) and \glsxtrshort{ure} (b,d) of the analytic gradients relative to the numerical gradients for the ground state of \ce{LiH} using a \glsxtrshort{sa-casscf} reference wave function. The top row (a,b) uses the tPBE0 function, and the bottom row (c,d) uses the MC23 functional.}
\end{figure}

\Cref{sfig:sa-num-vs-ana-curves} shows the common $\log$ of the \glsxtrshort{ue} and \glsxtrshort{ure} for the nuclear gradients of the ground and first singlet excited state of \ce{LiH} using SA-\glsxtrshort{mcpdft}. Both the \glsxtrshort{ue} and \glsxtrshort{ure} are stable with respect to the bond length, with the \glsxtrshort{ure} for both states increasing as the bond length increases due to the potential energy curves flattening out. Again, the errors are normally distributed and the histograms are shown in \cref{sfig:sa-err-distribution}. All \glsxtrshort{ue} are below \qty{1e-4}{\hartree\per\bohr} with the majority being on the order of \qty{1e-6}{\hartree\per\bohr}. The \glsxtrshort{ure} are slightly higher with the majority being on the order of \qty{1e-4}{}. 

\section{Supporting Figures}

\begin{figure}[H]
    \centering
    \includegraphics[width=\columnwidth]{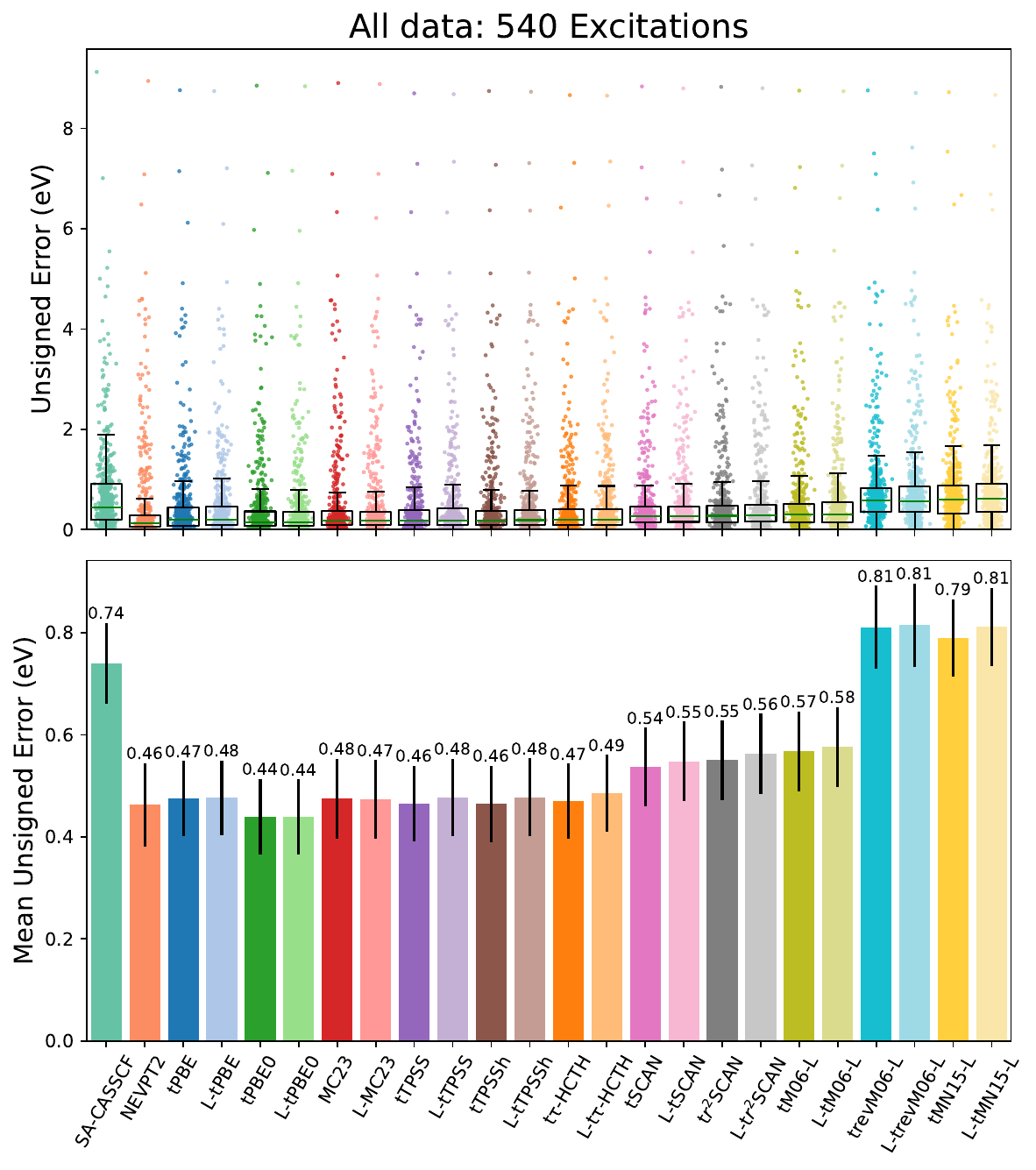}
    \caption{\label{sfig:questdb-all} Top: distribution of vertical excitation unsigned error and box and whisker plots over all 540 QUESTDB excitations for SA-CASSCF, NEVPT2, and 11 functionals for MC-PDFT and L-PDFT. Each dot represents a single unsigned error for the given method. The box represents the inter-quartile range and the horizontal green lines within each box indicates the median value. The whiskers extend $1.5\times$ the inter-quartile range. Bottom: bar plots of the mean unsigned error for each method on the 540 excitations with the black vertical lines indicating the \qty{95}{\percent} confidence intervals.}
\end{figure}

\begin{figure}[H]
    \centering
    \includegraphics[width=\columnwidth]{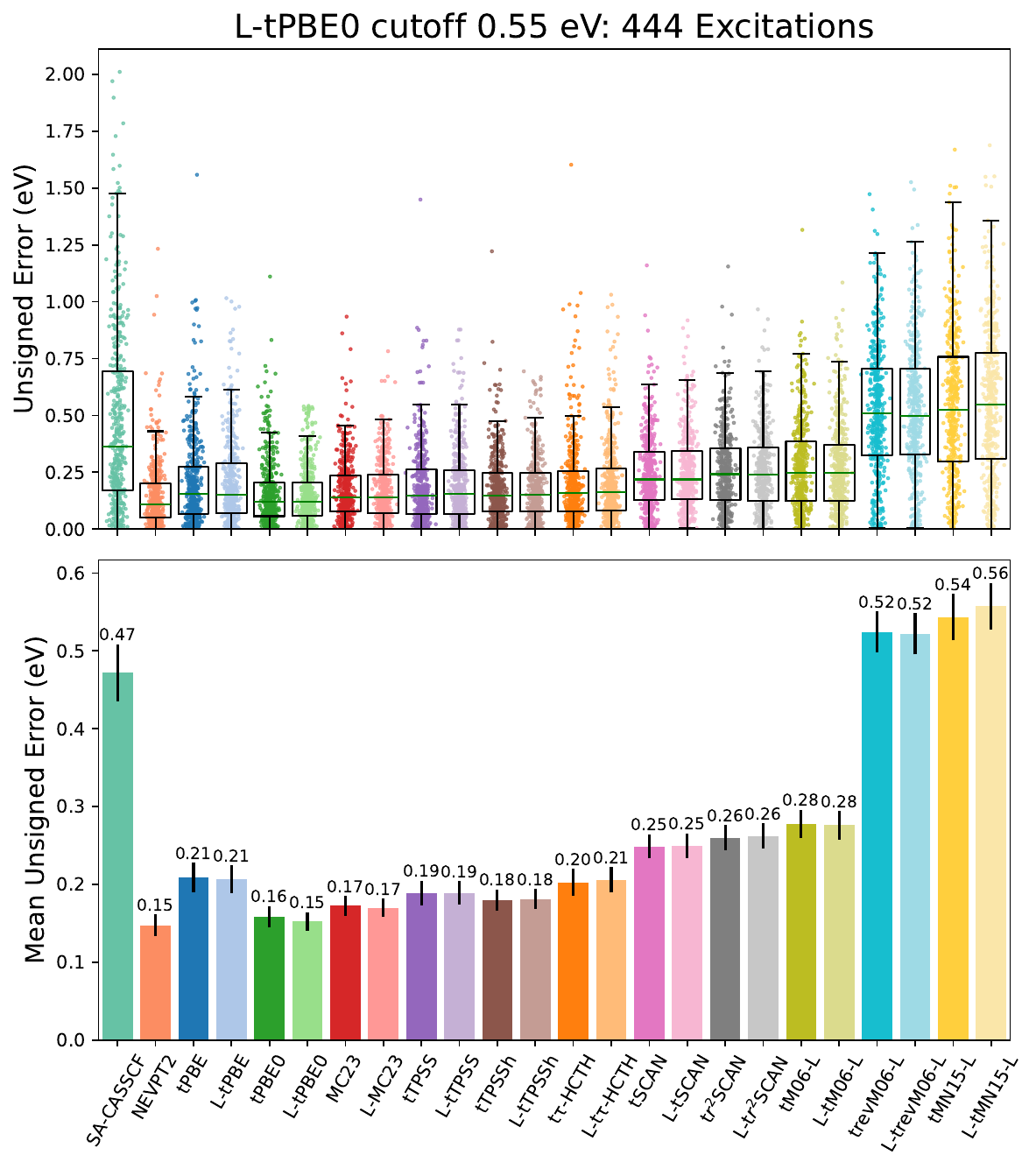}
    \caption{\label{sfig:questdb-ltpbe0-444} Top: distribution of vertical excitation unsigned error and box and whisker plots over 444 excitations with an L-tPBE0 error cutoff of \qty{0.55}{\eV} for SA-CASSCF, NEVPT2, and 11 functionals for MC-PDFT and L-PDFT. Each dot represents a single unsigned error for the given method. The box represents the inter-quartile range and the horizontal green lines within each box indicates the median value. The whiskers extend $1.5\times$ the inter-quartile range. Bottom: bar plots of the mean unsigned error for each method on the 44 excitations with the black vertical lines indicating the \qty{95}{\percent} confidence intervals.}
\end{figure}

\section{TD-DFT Excitations}

Here, we list the 359-excitations subset from the QUESTDB used to compare TD-DFT and MC-PDFT results.

\twocolumngrid
\setlist{nolistsep}
\begin{enumerate}[noitemsep]
    \item acetaldehyde-x1
\item acetaldehyde-x2
\item acetone-x1
\item acetone-x3
\item acetone-x6
\item acetone-x7
\item acetylene-x1
\item acetylene-x2
\item acetylene-x3
\item acetylene-x4
\item acetylene-x5
\item acrolein-x7
\item acrolein-x8
\item allyl-x1
\item allyl-x2
\item aza-naphthalene-x1
\item aza-naphthalene-x2
\item aza-naphthalene-x3
\item aza-naphthalene-x4
\item aza-naphthalene-x9
\item bef-x1
\item bef-x2
\item beh-x1
\item beh-x2
\item benzene-x1
\item benzene-x2
\item benzene-x3
\item benzene-x8
\item benzene-x9
\item benzene-x10
\item benzoquinone-x1
\item benzoquinone-x2
\item benzoquinone-x11
\item benzoquinone-x12
\item benzoquinone-x13
\item benzoquinone-x14
\item bh2-x1
\item butadiene-x1
\item butadiene-x7
\item butadiene-x8
\item carbon-monoxide-x1
\item carbon-monoxide-x2
\item carbon-monoxide-x3
\item carbon-monoxide-x4
\item carbon-monoxide-x7
\item carbon-monoxide-x8
\item carbon-monoxide-x9
\item carbon-monoxide-x10
\item carbon-monoxide-x11
\item carbonylfluoride-x1
\item carbonylfluoride-x2
\item ccl2-x1
\item ccl2-x2
\item ccl2-x3
\item ccl2-x4
\item cclf-x1
\item cf2-x2
\item ch3-x2
\item ch3-x3
\item cn-x1
\item cn-x2
\item co+-x1
\item co+-x2
\item cyanoacetylene-x1
\item cyanoacetylene-x2
\item cyanoacetylene-x3
\item cyanoacetylene-x4
\item cyanoformaldehyde-x1
\item cyanoformaldehyde-x2
\item cyanoformaldehyde-x3
\item cyanoformaldehyde-x4
\item cyanogen-x1
\item cyanogen-x2
\item cyanogen-x3
\item cyclopentadiene-x2
\item cyclopentadiene-x3
\item cyclopentadiene-x5
\item cyclopentadiene-x7
\item cyclopentadiene-x8
\item cyclopentadiene-x9
\item cyclopentadiene-x10
\item cyclopentadienethione-x1
\item cyclopentadienethione-x2
\item cyclopentadienethione-x4
\item cyclopentadienethione-x6
\item cyclopentadienethione-x7
\item cyclopentadienethione-x8
\item cyclopentadienone-x1
\item cyclopentadienone-x2
\item cyclopentadienone-x6
\item cyclopentadienone-x7
\item cyclopentadienone-x8
\item cyclopropene-x1
\item cyclopropene-x3
\item cyclopropene-x4
\item cyclopropenethione-x1
\item cyclopropenethione-x2
\item cyclopropenethione-x3
\item cyclopropenethione-x4
\item cyclopropenethione-x5
\item cyclopropenethione-x6
\item cyclopropenethione-x7
\item cyclopropenethione-x8
\item cyclopropenethione-x9
\item cyclopropenethione-x10
\item cyclopropenone-x1
\item cyclopropenone-x2
\item cyclopropenone-x3
\item cyclopropenone-x4
\item cyclopropenone-x5
\item cyclopropenone-x7
\item cyclopropenone-x8
\item cyclopropenone-x9
\item cyclopropenone-x10
\item cyclopropenone-x11
\item diacetylene-x1
\item diacetylene-x2
\item diacetylene-x3
\item diacetylene-x4
\item diazirine-x1
\item diazirine-x2
\item diazirine-x3
\item diazirine-x5
\item diazirine-x6
\item diazirine-x7
\item diazirine-x8
\item diazomethane-x1
\item diazomethane-x2
\item diazomethane-x3
\item diazomethane-x4
\item diazomethane-x5
\item diazomethane-x6
\item diazomethane-x7
\item difluorodiazirine-x1
\item difluorodiazirine-x2
\item difluorodiazirine-x3
\item difluorodiazirine-x4
\item difluorodiazirine-x5
\item difluorodiazirine-x6
\item dinitrogen-x1
\item dinitrogen-x2
\item dinitrogen-x3
\item dinitrogen-x5
\item dinitrogen-x8
\item dinitrogen-x9
\item dinitrogen-x10
\item dinitrogen-x11
\item ethylene-x1
\item ethylene-x3
\item ethylene-x5
\item ethylene-x6
\item ethylene-x7
\item f2bo-x1
\item f2bo-x2
\item f2bs-x1
\item f2bs-x2
\item formaldehyde-x1
\item formaldehyde-x2
\item formaldehyde-x4
\item formaldehyde-x7
\item formaldehyde-x9
\item formaldehyde-x10
\item formaldehyde-x11
\item formaldehyde-x13
\item formamide-x1
\item formamide-x5
\item formamide-x6
\item formylfluoride-x1
\item formylfluoride-x2
\item furan-x1
\item furan-x2
\item furan-x3
\item furan-x6
\item furan-x7
\item furan-x8
\item furan-x9
\item glyoxal-x1
\item glyoxal-x2
\item glyoxal-x4
\item glyoxal-x5
\item glyoxal-x6
\item glyoxal-x7
\item glyoxal-x8
\item glyoxal-x9
\item h2bo-x1
\item h2bo-x2
\item h2po-x1
\item h2po-x2
\item h2ps-x1
\item h2ps-x2
\item hccl-x1
\item hcf-x1
\item hco-x1
\item hcp-x1
\item hcp-x2
\item hcp-x3
\item hcp-x4
\item hexatriene-x1
\item hexatriene-x3
\item hexatriene-x4
\item hexatriene-x5
\item hexatriene-x6
\item hoc-x1
\item hpo-x1
\item hps-x1
\item hsif-x1
\item hydrogen-chloride-x1
\item hydrogen-sulfide-x1
\item hydrogen-sulfide-x2
\item hydrogen-sulfide-x3
\item hydrogen-sulfide-x4
\item imidazole-x1
\item imidazole-x2
\item imidazole-x3
\item imidazole-x5
\item imidazole-x6
\item imidazole-x7
\item imidazole-x8
\item isobutene-x2
\item isobutene-x3
\item ketene-x1
\item ketene-x2
\item ketene-x5
\item ketene-x6
\item ketene-x7
\item maleimide-x2
\item maleimide-x3
\item maleimide-x7
\item maleimide-x8
\item methanimine-x1
\item methanimine-x2
\item methylenecyclopropene-x1
\item methylenecyclopropene-x2
\item methylenecyclopropene-x3
\item methylenecyclopropene-x5
\item methylenecyclopropene-x6
\item naphthalene-x1
\item naphthalene-x2
\item naphthalene-x3
\item naphthalene-x4
\item nco-x1
\item nh2-x1
\item nitromethyl-x1
\item nitromethyl-x2
\item nitromethyl-x4
\item nitrosomethane-x1
\item nitrosomethane-x4
\item nitrosomethane-x5
\item octatetraene-x1
\item oh-x1
\item oh-x2
\item ph2-x1
\item propynal-x1
\item propynal-x2
\item propynal-x3
\item propynal-x4
\item pyrazine-x1
\item pyrazine-x2
\item pyrazine-x3
\item pyrazine-x4
\item pyrazine-x6
\item pyrazine-x7
\item pyrazine-x14
\item pyrazine-x15
\item pyrazine-x16
\item pyrazine-x17
\item pyrazine-x18
\item pyrazine-x19
\item pyridazine-x4
\item pyridazine-x6
\item pyridazine-x8
\item pyridazine-x9
\item pyridazine-x11
\item pyridine-x2
\item pyridine-x4
\item pyridine-x6
\item pyridine-x7
\item pyridine-x10
\item pyridine-x12
\item pyridine-x13
\item pyridine-x15
\item pyrimidine-x1
\item pyrimidine-x2
\item pyrimidine-x3
\item pyrimidine-x7
\item pyrimidine-x8
\item pyrimidine-x10
\item pyrimidine-x11
\item pyrrole-x1
\item pyrrole-x2
\item pyrrole-x3
\item pyrrole-x4
\item pyrrole-x5
\item pyrrole-x7
\item pyrrole-x8
\item pyrrole-x9
\item pyrrole-x10
\item sicl2-x1
\item sicl2-x2
\item silylidene-x1
\item silylidene-x2
\item streptocyanine-c1-x1
\item streptocyanine-c1-x2
\item streptocyanine-c3-x1
\item streptocyanine-c3-x2
\item streptocyanine-c5-x1
\item streptocyanine-c5-x2
\item tetrazine-x1
\item tetrazine-x2
\item tetrazine-x4
\item tetrazine-x5
\item tetrazine-x6
\item tetrazine-x7
\item tetrazine-x11
\item tetrazine-x12
\item tetrazine-x13
\item tetrazine-x15
\item tetrazine-x16
\item tetrazine-x17
\item thioacetone-x1
\item thioacetone-x2
\item thioacetone-x3
\item thioacetone-x6
\item thioacetone-x7
\item thioacrolein-x1
\item thioacrolein-x2
\item thioformaldehyde-x1
\item thioformaldehyde-x2
\item thioformaldehyde-x3
\item thioformaldehyde-x4
\item thioformaldehyde-x5
\item thioformaldehyde-x6
\item thiophene-x1
\item thiophene-x2
\item thiophene-x3
\item thiophene-x4
\item thiophene-x9
\item thiophene-x10
\item thiophene-x11
\item thiophene-x12
\item thiopropynal-x1
\item thiopropynal-x2
\item triazine-x1
\item triazine-x2
\item triazine-x3
\item triazine-x4
\item triazine-x5
\item triazine-x9
\item triazine-x10
\item triazine-x11
\item triazine-x12
\item triazine-x13
\item triazine-x14
\item water-x1
\item water-x2
\item water-x3
\item water-x4
\item water-x5
\item water-x6
\end{enumerate}

\onecolumngrid

\twocolumngrid

\begin{table}[H]
    \centering
    \caption{Mean unsigned error (MUE) on the 359-excitation subset for MC-PDFT, L-PDFT, SA-CASSCF, and NEVPT2. }
    \begin{tabular}{l S[table-format=1.2]}
        \toprule\toprule
        Method & {MUE} \\
        \midrule
        SA-CASSCF & 0.47 \\
        NEVPT2         & 0.15 \\
        tPBE           & 0.15 \\
        L-tPBE         & 0.16 \\
        MC23           & 0.17 \\
        L-MC23         & 0.18 \\
        tTPSS          & 0.18 \\
        L-tTPSS        & 0.20 \\
        tTPSSh         & 0.18 \\
        L-tTPSSh       & 0.19 \\
        t$\tau$-HCTH   & 0.20 \\
        L-t$\tau$-HCTH & 0.21 \\
        tSCAN          & 0.25 \\
        L-tSCAN        & 0.26 \\
        t$r^2$SCAN     & 0.26 \\
        L-t$r^2$SCAN   & 0.27 \\
        tM06-L         & 0.28 \\
        L-tM06-L       & 0.28 \\
        trevM06-L      & 0.53 \\
        L-trevM06-L    & 0.53 \\
        tMN15-L        & 0.54 \\
        L-tMN15-L      & 0.57 \\
        \bottomrule\bottomrule
    \end{tabular}
\end{table}

\begin{table}[H]
    \centering
    \caption{Mean unsigned error (MUE) on the 359-excitation subset for TD-DFT.}
    \begin{tabular}{l S[table-format=1.2]}
        \toprule\toprule
        Method & {MUE} \\
        \midrule
        M06-SX & 0.20 \\
        $\omega$B97X-D & 0.20 \\ 
        BMK & 0.20 \\
        $\omega$B97X-V & 0.21 \\
        SOGGA11-X & 0.22 \\
        CAM-B3LYP & 0.23 \\
        $\omega$B97M-V & 0.24 \\
        M06-2X & 0.24 \\
        LRC-$\omega$PBEh & 0.25 \\
        LRC-$\omega$PBE  & 0.25 \\
        MPW1K & 0.27 \\
        rCAM-B3LYP & 0.27 \\
        $\omega$M05-D & 0.28 \\
        PW6B95 & 0.28 \\
        MN15 & 0.28 \\
        PBE0 & 0.29 \\
        HSEHJS & 0.29 \\
        BHHLYP & 0.30 \\
        MPW1PW91 & 0.30 \\
        PBE50 & 0.32 \\
        M11 & 0.32 \\
        revTPSSh & 0.33 \\
        MS2h & 0.33 \\
        B3LYP & 0.34 \\
        B97M-V & 0.35 \\
        SCAN0 & 0.35 \\
        TPSSh & 0.35 \\
        M06-L & 0.36 \\
        mBEEF & 0.38 \\
        MS2 & 0.39 \\
        MVSh & 0.40 \\
        MVS & 0.41 \\
        revTPSS & 0.41 \\
        SCAN & 0.41 \\
        revM06-L & 0.44 \\
        TPSS & 0.44 \\
        MN15-L & 0.46 \\
        SPW92 & 0.48 \\
        B97-D & 0.49 \\
        PBE & 0.52 \\
        MPW91 & 0.53 \\
        BLYP & 0.55 \\
        MN12-SX & 0.57 \\
        \bottomrule\bottomrule
    \end{tabular}
\end{table}

\onecolumngrid

\section{Optimized Geometries}

\subsection{\textit{s-trans}-butadiene}

\begin{table}[H]
\centering
\caption{SA(2)-tPBE0(4,4)/jul-cc-pVTZ \textit{s-trans}-butadiene equilibrium ground-state geometry (in \unit{\angstrom}). State 0 energy: \qty{-155.59671794926683}{\hartree}. State 1 energy: \qty{-155.34596660711296}{\hartree}}.
\begin{tabular}{l d d d}
C & 1.16562 & 0.00002 & -1.43507 \\
C & 0.02921 & -0.00001 & -0.72863 \\
C & -0.02921 & 0.00001 & 0.72863 \\
C & -1.16562 & -0.00002 & 1.43507 \\
H & -0.92602 & -0.00004 & -1.25045 \\
H & 1.16140 & 0.00001 & -2.51899 \\
H & 2.13583 & 0.00006 & -0.94671 \\
H & 0.92602 & 0.00004 & 1.25045 \\
H & -1.16140 & -0.00001 & 2.51899 \\
H & -2.13583 & -0.00006 & 0.94671 \\
\end{tabular}
\end{table}

\begin{table}[H]
\centering
\caption{SA(2)-tPBE0(4,4)/jul-cc-pVTZ \textit{s-trans}-butadiene equilibrium excited-state geometry (in \unit{\angstrom}). State 0 energy: \qty{-155.55919329314588}{\hartree}. State 1 energy: \qty{-155.38788381476905}{\hartree}}.
\begin{tabular}{l d d d}
C & 1.26724 & 0.00003 & -1.51465 \\
C & 0.01136 & -0.00001 & -0.70085 \\
C & -0.01136 & 0.00001 & 0.70085 \\
C & -1.26724 & -0.00003 & 1.51465 \\
H & -0.93019 & -0.00005 & -1.24066 \\
H & 1.21966 & 0.00002 & -2.59375 \\
H & 2.23497 & 0.00007 & -1.03120 \\
H & 0.93019 & 0.00005 & 1.24066 \\
H & -1.21966 & -0.00002 & 2.59375 \\
H & -2.23497 & -0.00007 & 1.03120 \\
\end{tabular}
\end{table}

\begin{table}[H]
\centering
\caption{SA(2)-MC23(4,4)/jul-cc-pVTZ \textit{s-trans}-butadiene equilibrium ground-state geometry (in \unit{\angstrom}). State 0 energy: \qty{-155.9140836156452}{\hartree}. State 1 energy: \qty{-155.6600651148039}{\hartree}}.
\begin{tabular}{l d d d}
C & 1.16001 & 0.00002 & -1.42829 \\
C & 0.02498 & -0.00001 & -0.72515 \\
C & -0.02498 & 0.00001 & 0.72515 \\
C & -1.16001 & -0.00002 & 1.42829 \\
H & -0.92823 & -0.00004 & -1.24117 \\
H & 1.15729 & 0.00001 & -2.50707 \\
H & 2.12204 & 0.00006 & -0.93441 \\
H & 0.92823 & 0.00004 & 1.24117 \\
H & -1.15729 & -0.00001 & 2.50707 \\
H & -2.12204 & -0.00006 & 0.93441 \\
\end{tabular}
\end{table}

\begin{table}[H]
\centering
\caption{SA(2)-MC23(4,4)/jul-cc-pVTZ \textit{s-trans}-butadiene equilibrium excited-state geometry (in \unit{\angstrom}). State 0 energy: \qty{-155.87671993021522}{\hartree}. State 1 energy: \qty{-155.70151412409373}{\hartree}}.
\begin{tabular}{l d d d}
C & 1.25900 & 0.00003 & -1.50606 \\
C & 0.00591 & -0.00001 & -0.69834 \\
C & -0.00591 & 0.00001 & 0.69834 \\
C & -1.25900 & -0.00003 & 1.50606 \\
H & -0.93508 & -0.00005 & -1.22984 \\
H & 1.21518 & 0.00002 & -2.58036 \\
H & 2.21847 & 0.00007 & -1.01646 \\
H & 0.93508 & 0.00005 & 1.22984 \\
H & -1.21518 & -0.00002 & 2.58036 \\
H & -2.21847 & -0.00007 & 1.01646 \\
\end{tabular}
\end{table}

\subsection{Benzophenone}

\begin{table}[H]
\centering
\caption{SA(3)-CASSCF(8,7)/cc-pVDZ benzophenone equilibrium ground-state geometry (in \unit{\angstrom}). State 0 energy: \qty{-573.0728801223825}{\hartree}. State 1 energy: \qty{-572.9320483096009}{\hartree}. State 2 energy: \qty{-572.8243794668251}{\hartree}.}
\begin{tabular}{l d d d}
C & -1.29538 & 0.17462 & -0.29053 \\
C & 0.00000 & 0.00000 & -1.02775 \\
O & 0.00000 & 0.00000 & -2.24119 \\
C & 1.29538 & -0.17462 & -0.29053 \\
C & 2.48344 & 0.27072 & -0.89098 \\
C & 3.70016 & 0.06434 & -0.27776 \\
C & 3.76928 & -0.61470 & 0.93765 \\
C & 2.60328 & -1.08117 & 1.53019 \\
C & 1.37771 & -0.86208 & 0.92676 \\
H & 2.42993 & 0.77885 & -1.84269 \\
H & 4.60605 & 0.42680 & -0.74542 \\
H & 4.72607 & -0.78074 & 1.41426 \\
H & 2.65020 & -1.61945 & 2.46759 \\
H & 0.47962 & -1.23706 & 1.39712 \\
C & -1.37771 & 0.86208 & 0.92676 \\
C & -2.60328 & 1.08117 & 1.53019 \\
C & -3.76928 & 0.61470 & 0.93765 \\
C & -3.70016 & -0.06434 & -0.27776 \\
C & -2.48344 & -0.27072 & -0.89098 \\
H & -0.47962 & 1.23706 & 1.39712 \\
H & -2.65020 & 1.61945 & 2.46759 \\
H & -4.72607 & 0.78074 & 1.41426 \\
H & -4.60605 & -0.42680 & -0.74542 \\
H & -2.42993 & -0.77885 & -1.84269 \\
\end{tabular}
\end{table}

\begin{table}[H]
\centering
\caption{SA(3)-CASSCF(8,7)/cc-pVDZ benzophenone equilibrium excited-state geometry (in \unit{\angstrom}). State 0 energy: \qty{-573.042193181727}{\hartree}. State 1 energy: \qty{-572.9610317202039}{\hartree}. State 2 energy: \qty{-572.8534147576689}{\hartree}.}
\begin{tabular}{l d d d}
C & -1.30158 & 0.18026 & -0.21882 \\
C & 0.00000 & 0.00000 & -0.81881 \\
O & 0.00000 & 0.00000 & -2.18749 \\
C & 1.30158 & -0.18026 & -0.21882 \\
C & 2.46513 & 0.07964 & -0.98306 \\
C & 3.72024 & -0.10131 & -0.44767 \\
C & 3.87605 & -0.55061 & 0.86096 \\
C & 2.74157 & -0.83021 & 1.62239 \\
C & 1.48026 & -0.65769 & 1.10099 \\
H & 2.36792 & 0.44165 & -1.99667 \\
H & 4.59173 & 0.11369 & -1.05201 \\
H & 4.86350 & -0.68970 & 1.27889 \\
H & 2.85023 & -1.20445 & 2.63194 \\
H & 0.62226 & -0.92560 & 1.69794 \\
C & -1.48026 & 0.65769 & 1.10099 \\
C & -2.74157 & 0.83021 & 1.62239 \\
C & -3.87605 & 0.55061 & 0.86096 \\
C & -3.72024 & 0.10131 & -0.44767 \\
C & -2.46513 & -0.07964 & -0.98306 \\
H & -0.62226 & 0.92560 & 1.69794 \\
H & -2.85023 & 1.20445 & 2.63194 \\
H & -4.86350 & 0.68970 & 1.27889 \\
H & -4.59173 & -0.11369 & -1.05201 \\
H & -2.36792 & -0.44165 & -1.99667 \\
\end{tabular}
\end{table}

\begin{table}[H]
\centering
\caption{SA(3)-tPBE(8,7)/cc-pVDZ benzophenone equilibrium ground-state geometry (in \unit{\angstrom}). State 0 energy: \qty{-575.8341303027564}{\hartree}. State 1 energy: \qty{-575.6952761415855}{\hartree}. State 2 energy: \qty{-575.6192437388476}{\hartree}.}
\begin{tabular}{l d d d}
C & -1.29990 & 0.15853 & -0.28478 \\
C & 0.00000 & 0.00000 & -1.04518 \\
O & 0.00000 & 0.00000 & -2.26270 \\
C & 1.29990 & -0.15853 & -0.28478 \\
C & 2.48761 & 0.25154 & -0.91683 \\
C & 3.72485 & 0.04444 & -0.30104 \\
C & 3.79189 & -0.60995 & 0.93954 \\
C & 2.62869 & -1.04756 & 1.56099 \\
C & 1.37635 & -0.81554 & 0.95475 \\
H & 2.41811 & 0.72944 & -1.90088 \\
H & 4.64507 & 0.38625 & -0.79268 \\
H & 4.76690 & -0.78097 & 1.41569 \\
H & 2.67730 & -1.57058 & 2.52510 \\
H & 0.46172 & -1.17093 & 1.44580 \\
C & -1.37635 & 0.81554 & 0.95475 \\
C & -2.62869 & 1.04756 & 1.56099 \\
C & -3.79189 & 0.60995 & 0.93954 \\
C & -3.72485 & -0.04444 & -0.30104 \\
C & -2.48761 & -0.25154 & -0.91683 \\
H & -0.46172 & 1.17093 & 1.44580 \\
H & -2.67730 & 1.57058 & 2.52510 \\
H & -4.76690 & 0.78097 & 1.41569 \\
H & -4.64507 & -0.38625 & -0.79268 \\
H & -2.41811 & -0.72944 & -1.90088 \\
\end{tabular}
\end{table}

\begin{table}[H]
\centering
\caption{SA(3)-tPBE(8,7)/cc-pVDZ benzophenone equilibrium excited-state geometry (in \unit{\angstrom}). State 0 energy: \qty{-575.8088254363394}{\hartree}. State 1 energy: \qty{-575.7148537218917}{\hartree}. State 2 energy: \qty{-575.6271639320847}{\hartree}.}
\begin{tabular}{l d d d}
C & -1.29608 & 0.17524 & -0.23399 \\
C & 0.00000 & 0.00000 & -0.84513 \\
O & 0.00000 & 0.00000 & -2.19778 \\
C & 1.29608 & -0.17524 & -0.23399 \\
C & 2.46851 & 0.07881 & -1.00032 \\
C & 3.74024 & -0.10561 & -0.44380 \\
C & 3.88045 & -0.54179 & 0.87418 \\
C & 2.73872 & -0.81508 & 1.63686 \\
C & 1.45927 & -0.64736 & 1.09835 \\
H & 2.37266 & 0.44910 & -2.02901 \\
H & 4.62906 & 0.10467 & -1.05360 \\
H & 4.87953 & -0.67952 & 1.30808 \\
H & 2.84102 & -1.18653 & 2.66533 \\
H & 0.58193 & -0.92862 & 1.69111 \\
C & -1.45927 & 0.64736 & 1.09835 \\
C & -2.73872 & 0.81508 & 1.63686 \\
C & -3.88045 & 0.54179 & 0.87418 \\
C & -3.74024 & 0.10561 & -0.44380 \\
C & -2.46851 & -0.07881 & -1.00032 \\
H & -0.58193 & 0.92862 & 1.69111 \\
H & -2.84102 & 1.18653 & 2.66533 \\
H & -4.87953 & 0.67952 & 1.30808 \\
H & -4.62906 & -0.10467 & -1.05360 \\
H & -2.37266 & -0.44910 & -2.02901 \\
\end{tabular}
\end{table}

\begin{table}[H]
\centering
\caption{SA(3)-tPBE0(8,7)/cc-pVDZ benzophenone equilibrium ground-state geometry (in \unit{\angstrom}). State 0 energy: \qty{-575.142951025225}{\hartree}. State 1 energy: \qty{-575.0039152598074}{\hartree}. State 2 energy: \qty{-574.9196402972066}{\hartree}.}
\begin{tabular}{l d d d}
C & -1.29831 & 0.16424 & -0.28785 \\
C & 0.00000 & 0.00000 & -1.04129 \\
O & 0.00000 & 0.00000 & -2.25800 \\
C & 1.29831 & -0.16424 & -0.28785 \\
C & 2.48698 & 0.25666 & -0.91006 \\
C & 3.71814 & 0.05022 & -0.29329 \\
C & 3.78459 & -0.61046 & 0.94019 \\
C & 2.61945 & -1.05696 & 1.55229 \\
C & 1.37566 & -0.83018 & 0.94544 \\
H & 2.42331 & 0.74411 & -1.88505 \\
H & 4.63498 & 0.39823 & -0.77750 \\
H & 4.75409 & -0.77931 & 1.41811 \\
H & 2.66695 & -1.58413 & 2.50937 \\
H & 0.46514 & -1.19199 & 1.42971 \\
C & -1.37566 & 0.83018 & 0.94544 \\
C & -2.61945 & 1.05696 & 1.55229 \\
C & -3.78459 & 0.61046 & 0.94019 \\
C & -3.71814 & -0.05022 & -0.29329 \\
C & -2.48698 & -0.25666 & -0.91006 \\
H & -0.46514 & 1.19199 & 1.42971 \\
H & -2.66695 & 1.58413 & 2.50937 \\
H & -4.75409 & 0.77931 & 1.41811 \\
H & -4.63498 & -0.39823 & -0.77750 \\
H & -2.42331 & -0.74411 & -1.88505 \\
\end{tabular}
\end{table}

\begin{table}[H]
\centering
\caption{SA(3)-tPBE0(8,7)/cc-pVDZ benzophenone equilibrium excited-state geometry (in \unit{\angstrom}). State 0 energy: \qty{-575.1162529945068}{\hartree}. State 1 energy: \qty{-575.0255482756678}{\hartree}. State 2 energy: \qty{-574.9326933053055}{\hartree}.}
\begin{tabular}{l d d d}
C & -1.29757 & 0.17657 & -0.22909 \\
C & 0.00000 & 0.00000 & -0.83679 \\
O & 0.00000 & 0.00000 & -2.19431 \\
C & 1.29757 & -0.17657 & -0.22909 \\
C & 2.46727 & 0.07833 & -0.99571 \\
C & 3.73482 & -0.10536 & -0.44566 \\
C & 3.88005 & -0.54457 & 0.86966 \\
C & 2.74110 & -0.81881 & 1.63299 \\
C & 1.46608 & -0.64950 & 1.10010 \\
H & 2.37058 & 0.44562 & -2.02063 \\
H & 4.61873 & 0.10568 & -1.05465 \\
H & 4.87638 & -0.68282 & 1.29852 \\
H & 2.84605 & -1.19056 & 2.65651 \\
H & 0.59442 & -0.92618 & 1.69522 \\
C & -1.46608 & 0.64950 & 1.10010 \\
C & -2.74110 & 0.81881 & 1.63299 \\
C & -3.88005 & 0.54457 & 0.86966 \\
C & -3.73482 & 0.10536 & -0.44566 \\
C & -2.46727 & -0.07833 & -0.99571 \\
H & -0.59442 & 0.92618 & 1.69522 \\
H & -2.84605 & 1.19056 & 2.65651 \\
H & -4.87638 & 0.68282 & 1.29852 \\
H & -4.61873 & -0.10568 & -1.05465 \\
H & -2.37058 & -0.44562 & -2.02063 \\
\end{tabular}
\end{table}

\begin{table}[H]
\centering
\caption{SA(3)-MC23(8,7)/cc-pVDZ benzophenone equilibrium ground-state geometry (in \unit{\angstrom}). State 0 energy: \qty{-576.2437502524999}{\hartree}. State 1 energy: \qty{-576.1031903807286}{\hartree}. State 2 energy: \qty{-576.0165139431892}{\hartree}.}
\begin{tabular}{l d d d}
C & -1.29027 & 0.15918 & -0.28639 \\
C & 0.00000 & 0.00000 & -1.03801 \\
O & 0.00000 & 0.00000 & -2.25428 \\
C & 1.29027 & -0.15918 & -0.28639 \\
C & 2.47530 & 0.24157 & -0.91835 \\
C & 3.70226 & 0.03871 & -0.30140 \\
C & 3.76565 & -0.59881 & 0.93963 \\
C & 2.60106 & -1.02762 & 1.55981 \\
C & 1.36260 & -0.80452 & 0.95381 \\
H & 2.40889 & 0.70946 & -1.89837 \\
H & 4.61718 & 0.37067 & -0.79067 \\
H & 4.73075 & -0.76491 & 1.41753 \\
H & 2.64761 & -1.53742 & 2.52136 \\
H & 0.45232 & -1.15183 & 1.44092 \\
C & -1.36260 & 0.80452 & 0.95381 \\
C & -2.60106 & 1.02762 & 1.55981 \\
C & -3.76565 & 0.59881 & 0.93963 \\
C & -3.70226 & -0.03871 & -0.30140 \\
C & -2.47530 & -0.24157 & -0.91835 \\
H & -0.45232 & 1.15183 & 1.44092 \\
H & -2.64761 & 1.53742 & 2.52136 \\
H & -4.73075 & 0.76491 & 1.41753 \\
H & -4.61718 & -0.37067 & -0.79067 \\
H & -2.40889 & -0.70946 & -1.89837 \\
\end{tabular}
\end{table}

\begin{table}[H]
\centering
\caption{SA(3)-MC23(8,7)/cc-pVDZ benzophenone equilibrium excited-state geometry (in \unit{\angstrom}). State 0 energy: \qty{-576.2169072367556}{\hartree}. State 1 energy: \qty{-576.1250670452861}{\hartree}. State 2 energy: \qty{-576.0297439260203}{\hartree}.}
\begin{tabular}{l d d d}
C & -1.29127 & 0.17395 & -0.22531 \\
C & 0.00000 & 0.00000 & -0.83141 \\
O & 0.00000 & 0.00000 & -2.18623 \\
C & 1.29127 & -0.17395 & -0.22531 \\
C & 2.45454 & 0.07341 & -0.99523 \\
C & 3.71834 & -0.10777 & -0.44705 \\
C & 3.86350 & -0.53754 & 0.86768 \\
C & 2.72802 & -0.80686 & 1.63195 \\
C & 1.45661 & -0.63977 & 1.10192 \\
H & 2.35395 & 0.43276 & -2.01873 \\
H & 4.59863 & 0.09716 & -1.05557 \\
H & 4.85619 & -0.67372 & 1.29477 \\
H & 2.83393 & -1.17337 & 2.65262 \\
H & 0.58674 & -0.91533 & 1.69349 \\
C & -1.45661 & 0.63977 & 1.10192 \\
C & -2.72802 & 0.80686 & 1.63195 \\
C & -3.86350 & 0.53754 & 0.86768 \\
C & -3.71834 & 0.10777 & -0.44705 \\
C & -2.45454 & -0.07341 & -0.99523 \\
H & -0.58674 & 0.91533 & 1.69349 \\
H & -2.83393 & 1.17337 & 2.65262 \\
H & -4.85619 & 0.67372 & 1.29477 \\
H & -4.59863 & -0.09716 & -1.05557 \\
H & -2.35395 & -0.43276 & -2.01873 \\
\end{tabular}
\end{table}